\documentclass[lettersize,journal]{IEEEtran}
\PassOptionsToPackage{ruled,linesnumbered}{algorithm2e}
\usepackage[T1]{fontenc}
\usepackage[latin9]{inputenc}
\usepackage{verbatim}
\usepackage{refstyle}
\usepackage{units}
\usepackage{algorithm2e}
\usepackage{dsfont}
\usepackage{amsmath}
\usepackage{amsthm}
\usepackage{amssymb}
\usepackage{graphicx}
\usepackage{mleftright}
\mleftright

\makeatletter

\AtBeginDocument{\providecommand\appref[1]{\ref{app:#1}}}
\AtBeginDocument{\providecommand\algref[1]{\ref{alg:#1}}}
\AtBeginDocument{\providecommand\figref[1]{\ref{fig:#1}}}
\AtBeginDocument{\providecommand\thmref[1]{\ref{thm:#1}}}
\AtBeginDocument{\providecommand\subsecref[1]{\ref{subsec:#1}}}
\AtBeginDocument{\providecommand\subfigref[1]{\ref{subfig:#1}}}
\AtBeginDocument{\providecommand\lemref[1]{\ref{lem:#1}}}
\RS@ifundefined{subsecref}
{\newref{subsec}{name = \RSsectxt}}
{}
\RS@ifundefined{thmref}
{\def\RSthmtxt{theorem~}\newref{thm}{name = \RSthmtxt}}
{}
\RS@ifundefined{lemref}
{\def\RSlemtxt{lemma~}\newref{lem}{name = \RSlemtxt}}
{}

\theoremstyle{plain}
\newtheorem{thm}{\protect\theoremname}
\theoremstyle{remark}
\newtheorem{rem}{\protect\remarkname}
\theoremstyle{plain}
\newtheorem{lem}{\protect\lemmaname}
\usepackage{mathtools,empheq}
\usepackage{cite}
\usepackage{caption}
\usepackage{refstyle}
\usepackage[subrefformat=parens,labelformat=parens]{subfig}
\newcommand{\herm}{^{\mathsf{H}}}
\newcommand{\trans}{^{\mathsf{T}}}
\usepackage{amsmath}
\DeclareMathOperator{\Tr}{Tr}
\DeclareMathOperator{\st}{s.t.}
\DeclareMathOperator{\diag}{diag}

\DeclareMathOperator{\vect}{vec}
\DeclareMathOperator{\vecd}{vec_{d}}
\DeclareMathOperator{\argmax}{argmax}
\DeclareMathOperator{\argmin}{argmin}
\allowdisplaybreaks[1]
\usepackage{color}
\usepackage{nicefrac, xfrac}
\newref{fig}{refcmd = Fig.~\ref{#1}}
\newref{subfig}{refcmd = Fig.~\subref*{#1}}
\newref{alg}{refcmd = \textbf{Algorithm~\ref{#1}}}
\newref{app}{refcmd = Appendix~\ref{#1}}

\newref{subsec}{refcmd = Section~\ref{#1}}
\renewcommand*{\thmref}[1]{\textbf{Theorem~\ref{thm:#1}}}

\renewcommand*{\lemref}[1]{\textbf{Lemma~\ref{lem:#1}}}
\AtBeginDocument{\providecommand\eqref[1]{(\ref{eq:#1})}}

\@ifundefined{showcaptionsetup}{}{%
	\PassOptionsToPackage{caption=false}{subfig}}
\usepackage{subfig}
\makeatother

\usepackage{babel}
\providecommand{\lemmaname}{Lemma}
\providecommand{\remarkname}{Remark}
\providecommand{\theoremname}{Theorem}
\begin{document}
	\title{On the Joint Beamforming Design for Large-scale Downlink RIS-assisted Multiuser
		MIMO Systems}
	\author{Eduard E. Bahingayi, Nemanja~Stefan~Perovi\'c,
		\IEEEmembership{Member, IEEE}, Le-Nam Tran, \IEEEmembership{Senior Member, IEEE}\thanks{E. E. Bahingayi and L.-N. Tran are with the School of Electrical and
			Electronic Engineering, University College Dublin, Belfield, Dublin
			4, D04~V1W8, Ireland. Email: \{eduard.bahingayi,nam.tran\}@ucd.ie}\thanks{N.~S.~Perovi\'c is with Universit\'e Paris-Saclay, CNRS, CentraleSup\'elec,
			Laboratoire des Signaux et Syst\`emes, 3 Rue Joliot-Curie, 91192
			Gif-sur-Yvette, France. Email: nemanja-stefan.perovic@centralesupelec.fr}}
	\maketitle
	\begin{abstract}
		Reconfigurable intelligent surfaces (RISs) have huge potential to improve spectral and energy efficiency in future wireless systems at a minimal cost. However, early prototype results indicate that deploying hundreds or thousands of reflective elements is necessary for significant performance gains. Motivated by this, our study focuses on \emph{large-scale } RIS-assisted multi-user (MU) multiple-input multiple-output (MIMO) systems. In this context, we propose an efficient algorithm to jointly design the precoders at the base station (BS) and the phase shifts at the RIS to maximize the weighted sum rate (WSR). In particular, leveraging an equivalent lower-dimensional reformulation of the WSR maximization problem, we derive a closed-form solution to optimize the precoders using the successive convex approximation (SCA) framework. While the equivalent reformulation proves to be efficient for the precoder optimization, we offer numerical insights into why the original formulation of the WSR optimization problem is better suited for the phase shift optimization. Subsequently, we develop a scaled projected gradient method (SPGM) and a novel line search procedure to optimize RIS phase shifts. Notably, we show that the complexity of the proposed method \emph{scales linearly with the number of BS antennas and RIS reflective elements}. Extensive numerical experiments demonstrate that the proposed algorithm significantly reduces both time and computational complexity while achieving higher WSR compared to baseline algorithms.
	\end{abstract}
	

	\section{INTRODUCTION}
	Reconfigurable Intelligent Surfaces (RISs) have emerged as a disruptive technology for future wireless communications, gaining significant attention from both academia and industry \cite{strinati2021reconfigurable,pan2021reconfigurable,basharat2021reconfigurable}.	By employing programmable surfaces capable of dynamically manipulating electromagnetic waves, RISs can intelligently control the wireless propagation environment, thereby offering a wide range of advantages \cite{strinati2021reconfigurable,pan2021reconfigurable,basharat2021reconfigurable,kisseleff2020reconfigurable,wu2019towards}. One of the main advantages of RISs is their capability to improve link quality and signal coverage by establishing a virtual line-of-sight (LOS) channel between the base station (BS) and user equipment (UE), thereby reducing interference and boosting spectral efficiency. Additionally, RISs support secure communication and facilitate wireless information and power transfer \cite{basharat2021reconfigurable,pan2021reconfigurable,wu2019towards,hassouna2023survey}. All the benefits listed above highlight the potential of RISs to be a driving force in the evolution of future wireless networks.
	
	Our paper concentrates on the joint design of transmit precoder and phase shifts for large-scale RIS-aided systems, consisting of a large array of transmit antennas at the BS coupled with a large number of passive reflective elements at the RIS. This is motivated by the fact that the global deployment of fifth-generation (5G) wireless cellular networks is ramping up \cite{GMSA}, driven by the pivotal role of massive multiple-input multiple-output (MIMO) communications -- a cornerstone of 5G technology expected to persist into future wireless networks \cite{Larsson,rappaport2013millimeter,bertenyi20215g}. Also, early experimental testbeds of RIS have shown that a large number of passive reflective elements is required, ranging from hundreds to a few thousand \cite{dai2020reconfigurable,pei2021ris,wan2021reconfigurable}, so that this technology is capable of providing meaningful performance gain. In other words, beyond 5G wireless networks will feature an architecture with a large number of active and passive antenna elements \cite{pan2021reconfigurable}, which naturally calls for efficient algorithms for the joint optimization	of active and passive beamforming. Despite its importance, the domain of large-scale RIS-aided systems has been relatively under-explored in the existing research, presenting a crucial gap that this paper aims to address.
	
Since numerous studies have investigated RIS-aided wireless networks for various performance metrics, we attempt to provide a comprehensive (but by no means exhaustive) summary of the significant ones, especially those related to the considered problem. Concerning the RIS-aided single-user (SU) multiple-input single-output (MISO) systems, in \cite{wu2019intelligent}, the authors tackled the power minimization problem using the semidefinite relaxation (SDR) method despite its inherent high complexity. In related studies \cite{yu2019miso,yu2020optimal, JC2021Chen}, the authors focused on the rate maximization problem. Specifically, in \cite{yu2019miso}, a manifold optimization (MO)-based algorithm was employed for phase shift optimization, attaining superior performance and reduced complexity compared to the SDR-based method. An optimal solution based on the branch-and-bound (BnB) method for maximizing the achievable rate was proposed in \cite{yu2020optimal}. Additionally, the cross-entropy (CE) framework was proposed in \cite{JC2021Chen} to design RIS phase shifts with low resolution. Regarding the rate maximization problem for RIS-aided SU-MIMO systems, the authors in \cite{zhang2020capacity} derived a closed-form solution for optimizing transmit covariance and RIS phase shifts via an alternating optimization (AO) approach. This problem was further studied in \cite{perovic2021Ratemax} using a projected gradient approach, which showed significant improvements in achievable rates. The work of \cite{bahingayi2022low} considered millimeter-wave SU massive MIMO systems and proposed an AO approach to design passive and active beamforming matrices.

More recent research has been extended to RIS-aided multi-user (MU)-MIMO systems. The fundamental problem of transmit power minimization problem for MU-MISO downlink transmission was studied in \cite{Kumar2023SPMin}, and a successive convex approximation (SCA) method was proposed, which demonstrated significant improvements compared to \cite{wu2019intelligent}. Guo \emph{et al.} \cite{guo2020weighted} maximized the weighted sum rate (WSR) of the RIS-aided MU-MISO downlink system using a fractional programming (FP)-based framework. In the context of multigroup multicasting MISO systems, the study in \cite{zhou2020intelligent} proposed two algorithms based on the majorization-minorization (MM) technique to jointly optimize transmit and passive beamforming for the WSR maximization problem. The same problem was further investigated in \cite{farooq2022achievable} using the first-order method based on alternating projected gradient (APG). To the best of our knowledge, Perovi\'c, \emph{et al.} in \cite{perovic2022maximum} made the first attempt towards characterizing the capacity region of RIS-aided MIMO broadcast (BC) systems by the APG method, exploiting the duality between BC and multiple access channel (MAC). More recently, this problem was invested in \cite{DuongBC2024} for RISs with finite resolution. Furthermore, in  \cite{allu2023robust}, an AO-based algorithm was proposed to solve a sum mean-square-error (MSE) minimization problem for active RIS-aided MU-MIMO cognitive radio networks. In \cite{pan2020multicell}, the authors solved the sum-rate maximization problem for the RIS-aided MU-MIMO system by developing the WMMSE-based MM and MO approaches to update the RIS phase shifts. This study was extended to a double-RIS-aided MU-MIMO system in \cite{zhang2023double} to solve the MSE minimization problem.
In this paper, we address the WSR maximization problem for the {\emph{large-scale} RIS-aided MU-MIMO downlink system with linear precoding. To the best of our knowledge,  this problem has not been comprehensively addressed, if not at all. Most previous studies on RIS-aided MU-MIMO systems (e.g., \cite{guo2020weighted,allu2023robust,pan2020multicell,zhang2023double}) focus on scenarios with a relatively small number of BS antennas and RIS elements. The proposed solutions in these works often exhibit computational complexities that scale  \emph{cubically} with the number of BS antennas and/or RIS elements, rendering them impractical for large-scale RIS-aided MU-MIMO systems. We remark that the work in \cite{guo2020weighted} was dedicated to the RIS-aided MU-MISO systems, whereby each user has a single antenna. In contrast, we aim to develop \emph{linear-complexity} solutions for \emph{large-scale} RIS-aided MU-MIMO systems, i.e., multiple-antenna users. Our contributions are summarized as follows:
\begin{enumerate}
    \item We propose a numerically efficient algorithm based on AO for maximizing the WSR for large-scale RIS-aided MU-MIMO systems, in which the BS is constrained by the total power budget and the RIS reflecting elements are subject to the unit modulus constraint. Specifically, the considered problem is decomposed into two subproblems: transmit precoder optimization and RIS phase shifts optimization, which are solved separately in an alternating manner. For optimizing the transmit precoders, we invoke an interesting result recently reported in \cite{zhao2023rethinking}, which states that the WSR maximization problem for MU-MIMO systems admits an equivalent unconstrained low-dimensional optimization problem. However, contrary to the weighted minimum mean squared error (WMMSE) approach utilized in \cite{zhao2023rethinking}, our approach employs the SCA technique to derive a closed-form solution for transmit precoder optimization.
   \item We provide numerical insights into why the equivalent formulation of the WSR maximization problem may not be efficient for RIS phase shift optimization. To this end, we empirically investigate the Lipschitz constant of the gradient of the RIS phase shifts for the original and equivalent formulations. Based on this observation, we propose a scaled projected gradient method (SPGM) algorithm to update the RIS phase shifts in each iteration of the proposed method. In this regard, we also propose a novel line-search method to find a proper step size in each step of the SPGM algorithm. Our extensive	numerical experiments demonstrate that the proposed SPGM algorithm converges faster than the conventional counterparts.
   \item We conduct a thorough complexity analysis of our proposed algorithm for optimizing the transmit precoder and the RIS phase shifts, showing that its complexity scales linearly with the number of transmit antennas and RIS reflective elements.
   \item We carry out extensive numerical simulations to evaluate the effectiveness of our proposed algorithm, resulting in significant reductions in both time and computational complexity. Moreover, our proposed algorithm consistently outperforms benchmark algorithms in achieving larger WSR.
\end{enumerate}
	
    The remainder of this paper is organized as follows: In Section \ref{sec:SystemModel}, we discuss both our system model and problem formulation. Section \ref{sec:PropSolution} presents the proposed algorithms.	In Section \ref{sec:SimResults}, we provide the simulation results to evaluate the performance of the proposed algorithm numerically. Finally, Section \ref{sec:Conclusion} offers some concluding remarks.
    
    \emph{Notation}: Throughout the paper, we adhere to the following notations: upper and lowercase boldface letters denote matrices and vectors, respectively. The $n$-th entry of $\mathbf{x}$ is denoted by $x_{n}$ and $\mathbf{x}_{n}$ denotes the $n$-th column of $\mathbf{X}$. We use $\mathbf{X}^{*}$, $\mathbf{X}\trans$, $\mathbf{X}\herm$, and $\mathbf{X}^{-1}$ to represent the conjugate, transpose, Hermitian, and inverse of $\mathbf{X}$, respectively. $\Tr(\mathbf{X})$ and $\det(\mathbf{X})$ denote the trace and determinant of
	$\mathbf{X}$, respectively. $\mathbf{X}\succeq0$ denotes that $\mathbf{X}$ is a positive semi-definite matrix. The operator $\diag(\mathbf{X_1},\cdots \mathbf{X_N})$ denotes a block diagonal matrix with  $\mathbf{X_1},\cdots \mathbf{X_N}$ as its diagonal blocks. If the diagonal blocks are scalars, this reduces to a standard diagonal matrix. The operator $\vecd(\mathbf{X})$ denotes the vector of the diagonal elements of $\mathbf{X}$. $\mathbf{\nabla}_{\mathbf{X}}f(\cdot)$ denotes the complex-valued gradient of $f$ with respect to (w.r.t)	$\mathbf{X}$ as defined in \cite{hjorungnes2011complex}, i.e., $\mathbf{\nabla}_{\mathbf{X}}f(\mathbf{X}) =\frac{1}{2}(\frac{\partial}{\partial\Re\{\mathbf{X}\}}f(\mathbf{X})+j\frac{\partial}{\partial\Im\{\mathbf{X}\}}f(\mathbf{X}))$. The operator $\Re\{\cdot\}$ extracts the real part of the argument, $|x|$ returns the absolute value of a complex number $x$, $\log(x)$ denotes the natural logarithm of $x$, and $\mathbb{E}\{\cdot\}$ denotes the expectation operator. The complex matrix space of size $m\times n$ is represented by $\mathbb{C}^{m\times n}$. Finally, $\left\Vert \mathbf{X}\right\Vert $ denotes the Frobenius norm of $\mathbf{X}$ (which is reduced to the Euclidean norm if the argument is a vector $\mathbf{x}$) and $\otimes$ denotes the Kronecker product.	
    \section{System Model and Problem Formulation\label{sec:SystemModel}}    
    \subsection{System Model}We consider a RIS-aided MU-MIMO downlink system illustrated in Fig. \ref{fig:SystemArch} which consists of a BS with $N_{t}$ antennas, a RIS with $N_{s}$ reflecting elements, and $K$ users each equipped with $N_{r}$ antennas.\footnote{To lighten our notation, we have assumed that each user has the same number of receive antennas. However, it is straightforward to adapt our mathematical framework to deal with varying numbers of antennas	per user.} The equivalent baseband channel of the BS-RIS, RIS-user $k$, and	direct BS-user $k$ links are denoted by $\mathbf{G}\ensuremath{\in\mathbb{C}^{N_{s}\times N_{t}}}$,	$\mathbf{U}_{k}\in\mathbb{C}^{N_{r}\times N_{s}}$, and $\mathbf{D}_{k}\in\mathbb{C}^{N_{r}\times N_{t}}$, respectively. Furthermore, the RIS phase shift matrix is denoted	by $\mathbf{T}(\boldsymbol{\theta})=\diag(\boldsymbol{\theta})$ where	$\boldsymbol{\theta} =\left[e^{j\varphi_{1}},e^{j\varphi_{2}},\cdots,e^{j\varphi_{N_{s}}}\right]\trans\in\mathbb{C}^{N_{s}\times1}$,	and $\varphi_{n}$ is the phase shift of the $n$-th element of the RIS. Thus, the composite channel between the BS and user $k$ is given by.
	\begin{align}
		\mathbf{H}_{k}(\boldsymbol{\theta}) & =\mathbf{D}_{k}+\mathbf{U}_{k}\mathbf{T}(\boldsymbol{\theta})\mathbf{G}.
	\end{align}
	Accordingly, the signal received by user $k$ is expressed as
	\begin{align}
		\mathbf{y}_{k} & =\mathbf{H}_{k}(\boldsymbol{\theta})\mathbf{W}_{k}\mathbf{s}_{k}+\mathbf{H}_{k}(\boldsymbol{\theta})\sum\nolimits_{j\neq k}^{K}\mathbf{W}_{j}\mathbf{s}_{j}+\mathbf{z}_{k},\label{eq:receivesig}
	\end{align}
	where $\mathbf{W}_{k}\in\mathbb{C}^{N_{t}\times N_{d}}$ is the linear precoder for user $k$, $\mathbf{s}_{k}\in\mathbb{C}^{N_{d}\times1}$ denotes the $N_{d}\leq N_{r}$ data streams transmitted to user $k$, which satisfies $\mathbf{\mathbb{E}}\left\{ \mathbf{s}_{k}\mathbf{s}_{k}\herm\right\} =\mathbf{I}$, and $\mathbf{z}_{k}\sim\mathcal{CN}(0,\sigma_{0}^{2}\mathbf{I})$ represents an independent and identically distributed additive white Gaussian noise, where $\sigma_{0}^{2}$ is the noise power. In this paper, we assume the availability of perfect channel state information (CSI) at the BS, as is commonly done in many previous studies. While this assumption is idealized, it is a standard approach in the literature (e.g., \cite{ zhang2020capacity, perovic2021Ratemax,bahingayi2022low, Kumar2023SPMin, guo2020weighted, zhou2020intelligent,farooq2022achievable,perovic2022maximum, pan2020multicell, DuongBC2024}) to characterize the fundamental performance limits of the system under consideration. Specifically, the assumption of perfect CSI allows us to gain valuable insights into the achievable sum rate of large-scale RIS-aided MU-MIMO systems. Despite the challenges in acquiring CSI related to RIS in practice due to the passive nature of the RIS, CSI estimation methods for RIS-aided MU-MIMO systems, such as those in \cite{chen2023channel,liu2020matrix,wei2021channel} are available and can be applied to our considered system model. Treating the multiuser interference in (\ref{eq:receivesig}) as Gaussian noise, an achievable rate for user $k$ is given by
	\begin{align}
		R_{k}(\mathbf{W},\boldsymbol{\theta}) & =\log\det\bigl(\sigma_{0}^{2}\mathbf{I}+\sum_{j=1}^{K}\nolimits\mathbf{H}_{k}(\boldsymbol{\theta})\mathbf{W}_{j}\mathbf{W}_{j}\herm\mathbf{H}_{k}(\boldsymbol{\theta})\herm\bigr)\nonumber \\
		&\hspace{-20pt}-\log\det\bigl(\sigma_{0}^{2}\mathbf{I}+\sum_{j\neq k}^{K}\nolimits\mathbf{H}_{k}(\boldsymbol{\theta})\mathbf{W}_{j}\mathbf{W}_{j}\herm\mathbf{H}_{k}(\boldsymbol{\theta})\herm\bigr).\label{eq:rate-userk-full}
	\end{align}
	Note that for mathematical convenience, we have used the natural logarithm in the above equation, and thus the achievable rate is expressed in	nat/s/Hz.
    \begin{figure}[!t]
        \centering{}\includegraphics[scale=0.30]{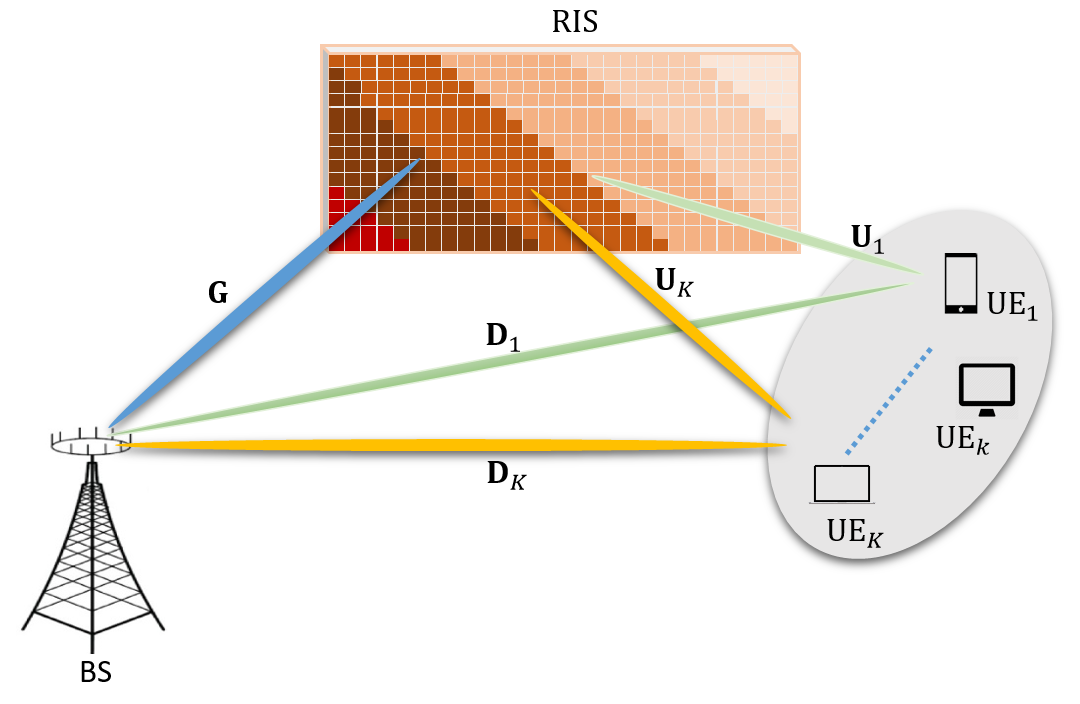}\caption{Illustration of a downlink RIS-aided MU-MIMO system}\label{fig:SystemArch}
    \end{figure}
    \subsection{Problem Formulation}
    In this paper, we are interested in the fundamental problem of WSR maximization by jointly optimizing the transmit precoder $\mathbf{W}$ and the RIS's phase shifts $\boldsymbol{\theta}$. Mathematically, the considered optimization problem is stated as
\begin{subequations}\label{eq:WSR}
\begin{IEEEeqnarray}{rCl}
&\max_{\mathbf{W},\boldsymbol{\theta}}  &\quad R(\mathbf{W, }\boldsymbol{\theta})\triangleq \sum\nolimits_{k=1}^{K}\omega_{k}R_{k}(\mathbf{W},\boldsymbol{\theta}),
\\*
({\mathcal{P}_1}) : \smash{\left\{
\IEEEstrut[9\jot]
\right.}
&{\rm s.t.}  &\quad |\theta_{n}|=1,\thinspace\forall n=1,\cdots,N_{s}, \label{eq:modulus}
\\*
&  & \quad\sum\nolimits_{k=1}^{K}{\rm Tr}(\mathbf{W}_{k}\mathbf{W}^{H}_{k})\leq P_{{\rm BS}}.\label{eq:SPC}
\end{IEEEeqnarray}
\end{subequations} 
	In the above formulation,  $\omega_{k}\thinspace(\sum_{k=1}^{K}{\omega_{k}=1})$ denotes the priority weight for user	$k$, which is a predetermined constant to achieve a specific scheduling goal, such as ensuring user fairness, $\eqref{eq:modulus}$ ensures that the RIS elements do not amplify the power of the incident signals, and $\eqref{eq:SPC}$ is the total power constraint at the BS. Since the objective function $(\mathcal{P}_{1})$	is non-convex, it is difficult to find the optimal solution. In fact, for a given $\boldsymbol{\theta}$, $(\mathcal{P}_{1})$ is proved to be NP-hard \cite{NPhard1}. As such, $(\mathcal{P}_{1})$ largely remains an open and challenging problem. Some attempts were made to solve $(\mathcal{P}_{1})$, but these approaches have critical drawbacks that are discussed next.\par
	Before introducing our proposed solutions in the subsequent section, we provide critical remarks regarding the existing solutions to $(\mathcal{P}_{1})$. We notice the complexities of WMMSE-based methods in \cite{pan2020multicell} for optimizing $\mathbf{W}$ and $\boldsymbol{\theta}$ are in the order of $\mathcal{O}(N_{t}^{3})$ and $\mathcal{O}(N_{s}^{3})$, which is obviously not  impractical for very large $N_{t}$ and $N_{s}$. To the best of our knowledge, the study in \cite{guo2020weighted} provides the state-of-the-art solution for the WSR maximization problem for MISO systems, with relatively low complexity. Specifically, the authors in \cite{guo2020weighted} developed two AO-based algorithms to jointly optimize $\mathbf{W}$ and $\boldsymbol{\theta}$ in $(\mathcal{P}_{1})$. In their first approach, they utilized the WMMSE and Riemannian conjugate gradient (RCG) methods to alternately optimize $\mathbf{W}$ and $\boldsymbol{\theta}$, respectively. However, this approach is notably computation-intensive. Specifically, the WMMSE algorithm requires the inversion of an $N_{t}$-dimensional matrix with the complexity of $\mathcal{O}(N_{t}^{3})$, and it involves an iterative bi-sectional search method to satisfy the transmit power constraint, which significantly slows convergence. To reduce the complexity of the first method, the authors applied the FP framework, leading to a block coordinate descent (BCD)-based algorithm to jointly optimize $\mathbf{W}$ and $\boldsymbol{\theta}$ in $(\mathcal{P}_{1})$. However, the per-iteration complexity of the BCD algorithm for optimizing $\mathbf{W}$ and $\boldsymbol{\theta}$, is still in the order of $\mathcal{O}(N_{t}^{2})$ and $\mathcal{O}(N_{s}^{2})$, respectively. It is obvious that the BCD algorithm is still not practically suitable for systems with large $N_{t}$ and $N_{s}$. To address this issue, we propose in the next section an efficient solution whose complexity \emph{scales linearly with the number of RIS elements, i.e., $\mathcal{O}(N_{s})$ and BS antennas, i.e., $\mathcal{O}(N_{t})$}. This is a massive complexity reduction compared to \cite{guo2020weighted}.
    	
    \section{Proposed Solution \label{sec:PropSolution}}
     First, it is noteworthy that the feasible sets of $\mathbf{W}$ and $\boldsymbol{\theta}$ in $(\mathcal{P}_{1})$ are decoupled, making AO a natural choice since it can find a stationary solution, which is an important result for non-convex problems. The AO approach has	been indeed a popular optimization tool in a majority of studies in RIS-related literature \cite{wu2019intelligent,yu2019miso, JC2021Chen,perovic2022maximum, farooq2022achievable,guo2020weighted}.	The fundamental concept behind the AO approach is to optimize one	variable while keeping other variables fixed. Thus, in this paper, we also adopt the AO framework to solve $(\mathcal{P}_{1})$. Specifically, let $(\mathbf{W}^{(\ell)},\boldsymbol{\theta}^{(\ell)})$ denote the value of $\mathbf{W}$ and $\boldsymbol{\theta}$ at the $\ell$-th iteration of the proposed AO algorithm. Then we update $\mathbf{W}$ and $\boldsymbol{\theta}$ as follows
$
    (\mathbf{W}^{(\ell)},\boldsymbol{\theta}^{(\ell)})\to (\mathbf{W}^{(\ell+1)},\boldsymbol{\theta}^{(\ell)}) \to (\mathbf{W}^{(\ell+1)},\boldsymbol{\theta}^{(\ell+1)})\to\cdots.
$
    In the subsequent subsections, we describe the novelty of our proposed AO-based method for the two optimization steps above to obtain $(\mathbf{W}^{(\ell+1)},\boldsymbol{\theta}^{(\ell+1)})$ from the current iterate $(\mathbf{W}^{(\ell)},\boldsymbol{\theta}^{(\ell)})$.
	
    \subsection{Transmit Beamforming Optimization: $\mathbf{W}$-update \label{subsec:Active-Beamforming}}
    When $\boldsymbol{\theta}$ is fixed, the sub-problem to optimize $\mathbf{W}$ becomes the classical WSR maximization problem for MIMO downlink channel, for which well-established solutions exist in the literature, primarily including the SCA \cite{ShiSCA1, KimSCA2, NamSCA3, NguyenSCA4} and the WMMSE \cite{shi2011iteratively, ChristensenWMMSE} algorithms. In these studies, the optimization methods are based directly on $(\mathcal{P}_{1})$, which involve large-sized matrices for large $N_{t}$, resulting in high computational complexity and slow convergence.
	
    In contrast to previous research, we consider an equivalent reformulation of $(\mathcal{P}_{1})$ that has a smaller dimension, which obviously facilitates the development of more efficient solutions. This is made possible by an interesting recent result reported in \cite{zhao2023rethinking}. Specifically, let $\mathbf{H}$ denote the matrix that includes all user's channel, i.e. $\mathbf{H}=[\mathbf{H}_{1}\trans,\mathbf{H}_{2}\trans,\ldots,\mathbf{H}_{K}\trans]\trans\in\mathbb{C}^{KN_{r}\times N_{t}}$, where for brevity, we have dropped the dependency of the user's channels on $\boldsymbol{\theta}$. Then, Proposition 5 in \cite{zhao2023rethinking} asserts that $(\mathcal{P}_{1})$ is equivalent to the following problem:   

\begin{IEEEeqnarray}{C}
    ({\mathcal{P}_2}): \quad \max_{\mathbf{F} }  \, \tilde{R}(\mathbf{F})\triangleq \sum\nolimits_{k=1}^{K}\omega_{k}\tilde{R}_{k}(\mathbf{F}),
\end{IEEEeqnarray}
where 
\begin{align}
		\tilde{R}_{k}(\mathbf{F}) & =\log\det\bigl(\mathbf{I}+\bar{\mathbf{H}}_{k}\mathbf{F}_{k}\mathbf{F}_{k}\herm\bar{\mathbf{H}}_{k}\herm
		\bigl(\sum\nolimits_{j\neq k}^{K}\bar{\mathbf{H}}_{k}\mathbf{F}_{j}\mathbf{F}_{j}\herm\bar{\mathbf{H}}_{k}\herm \nonumber \\
		&\hspace{50pt}+\frac{\sigma_{0}^{2}}{P_{{\rm BS}}}\sum\nolimits _{i=1}^{K}\Tr(\mathbf{F}_{i}\herm\bar{\mathbf{H}}\mathbf{F}_{i})\mathbf{I}\bigr)^{-1}\bigr),\label{eq:rate-equivalent}
\end{align}
	$\mathbf{F}=[\mathbf{F}_{1},\ldots\mathbf{F}_{K}]\in\mathbb{C}^{KN_{r}\times KN_{d}}$ is the new optimization variable, $\bar{\mathbf{H}}=\mathbf{H}\mathbf{H}\herm\in\mathbb{C}^{KN_{r}\times KN_{r}}$, and $\bar{\mathbf{H}}_{k}=\mathbf{H}_{k}\mathbf{H}^{H}\in\mathbb{C}^{N_{r}\times KN_{r}}$. Note that optimization over $\mathbf{W}$ in $(\mathcal{P}_{1})$ has been equivalently converted to optimization over $\mathbf{F}$ in $(\mathcal{P}_{2})$, which leads to two significant advantages. First, $(\mathcal{P}_{2})$ involves substantially fewer optimization variables than $(\mathcal{P}_{1})$ when $N_{t}$ is large, i.e. $K^{2}N_{r}N_{d}$ as opposed to $KN_{t}N_{d}$. Second, unlike $(\mathcal{P}_{1})$, $(\mathcal{P}_{2})$ is an unconstrained optimization problem, since the transmit power constraint is reflected in the objective. After $\mathbf{F}$ is found, then $\mathbf{W}$ is computed as $\mathbf{W}=\sqrt{\xi}\mathbf{H}\herm\mathbf{F}$, where $\xi=\frac{P_{BS}}{||\mathbf{H}\herm\mathbf{F}||^{2}}$ is power-normalization factor to ensure $\eqref{eq:SPC}$ is met. For the sake of brevity, we omit the proof of the equivalence between $(\mathcal{P}_{1})$ and $(\mathcal{P}_{2})$ and refer the interested readers to \cite{zhao2023rethinking} for the details.
 
    We now propose an efficient solution for solving $(\mathcal{P}_{2})$, which is different from \cite{zhao2023rethinking}. Specifically, in \cite{zhao2023rethinking}, the authors introduced reduced WMMSE (R-WMMSE) to solve $(\mathcal{P}_{2})$, employing three closed-form updates per iteration, similar to the conventional WMMSE algorithm in \cite{shi2011iteratively}. Thus, the R-WMMSE exhibits slow convergence as it has to perform three sequential updates during each iteration. Motivated by this drawback of the R-WMMSE algorithm, we propose an approach based on SCA to tackle $(\mathcal{P}_{2})$. In this regard, the main challenge in applying SCA to solve $(\mathcal{P}_{2})$ is to derive a proper bound for the objective function such that the resulting subproblem admits a closed form solution.
    
    To proceed, let us consider the following inequality for $\mathbf{X}_{k}$ and $\hat{\mathbf{X}}_{k}$ with size $p\times q$ and $\mathbf{Y}_{k}\succeq0$ and $\mathbf{\hat{\mathbf{Y}}}_{k}\succeq0$ with size $p\times p$ \cite{tam2016successive},
\begin{align}		\log\det(\mathbf{I}+\mathbf{X}_{k}\herm\mathbf{Y}_{k}^{-1}\mathbf{X}_{k}) & \geq\log\det(\mathbf{I}+\hat{\mathbf{X}}_{k}\herm\hat{\mathbf{Y}}_{k}^{-1}\hat{\mathbf{X}}_{k})\nonumber \\
		&\hspace{-60pt}-\Tr(\mathbf{\hat{\mathbf{X}}}_{k}\herm\hat{\mathbf{Y}}_{k}^{-1}\hat{\mathbf{X}}_{k})+2\Re\{\Tr(\hat{\mathbf{X}}_{k}\herm\hat{\mathbf{Y}}_{k}^{-1}\mathbf{X}_{k})\}\nonumber \\
		&\qquad-\Tr(\hat{\mathbf{A}}_{k}\herm(\mathbf{X}_{k}\mathbf{X}_{k}\herm+\mathbf{Y}_{k})),\label{eq:enequlityXY}
\end{align}
where $\hat{\mathbf{A}}_{k}\triangleq\mathbf{\hat{\mathbf{Y}}}_{k}^{-1}-(\mathbf{\hat{\mathbf{X}}}_{k}\mathbf{\hat{\mathbf{X}}}_{k}\herm+\mathbf{\hat{\mathbf{Y}}}_{k})^{-1}\succeq0$. A more elegant proof of the above inequality is presented in \appref{proof-key-inequality}. Note that equality occurs when $(\mathbf{X}_{k},\mathbf{Y}_{k})=(\hat{\mathbf{X}}_{k},\hat{\mathbf{Y}}_{k})$ and that the complex-value gradients on both sides w.r.t. $\mathbf{X}_{k}$ and $\mathbf{Y}_{k}$, respectively, are identical. Thus, (\ref{eq:enequlityXY}) can be used to obtain a tight concave lower bound approximation of $\tilde{R}_{k}(\mathbf{F})$ in the context of SCA as follows. Let $\mathbf{F}^{(n)}=[\mathbf{F}_{1}^{(n)},\mathbf{F}_{2}^{(n)},\ldots,\mathbf{F}_{K}^{(n)}]$ denote the value of $\mathbf{F}$ after $n$ iterations, $\mathbf{X}_{k}=\bar{\mathbf{H}}_{k}\mathbf{F}_{k}$, $\hat{\mathbf{X}}_{k}=\bar{\mathbf{H}}_{k}\mathbf{F}_{k}^{(n)}$, $\mathbf{Y}_{k}=\sum_{j\neq k}^{K}\bar{\mathbf{H}}_{k}\mathbf{F}_{j}\mathbf{F}_{j}\herm\bar{\mathbf{H}}_{k}\herm+\frac{\sigma_{0}^{2}}{P_{{\rm BS}}}\sum_{i=1}^{K}\Tr(\bar{\mathbf{H}}\mathbf{F}_{i}\mathbf{F}_{i}\herm)\mathbf{I}$ and $\hat{\mathbf{Y}}_{k}=\sum_{j\neq k}^{K}\bar{\mathbf{H}}_{k}\mathbf{F}_{j}^{(n)}(\mathbf{F}_{j}^{(n)})\herm\bar{\mathbf{H}}_{k}\herm+\frac{\sigma_{0}^{2}}{P_{{\rm BS}}}\sum_{i=1}^{K}\Tr(\bar{\mathbf{H}}\mathbf{F}_{i}^{(n)}(\mathbf{F}_{i}^{(n)})\herm)\mathbf{I}$. After a few simple algebraic manipulations, (\ref{eq:enequlityXY}) results in
        \begin{align}
		\tilde{R}_{k}(\mathbf{F})&\geq g_{k}(\mathbf{F})\triangleq  \log\det(\mathbf{I}+\hat{\mathbf{B}}_{k}\mathbf{\hat{\mathbf{X}}}_{k})-\Tr(\hat{\mathbf{B}}_{k}\mathbf{\hat{\mathbf{X}}}_{k})\nonumber \\
		&+2\Re\{\Tr(\hat{\mathbf{B}}_{k}\bar{\mathbf{H}}_{k}\mathbf{F}_{k})\}-\sum\nolimits _{j=1}^{K}\Tr(\mathbf{F}_{j}\herm\bar{\mathbf{H}}_{k}\herm\mathbf{A}_{k}\herm \bar{\mathbf{H}}_{k}\mathbf{F}_{j})\nonumber \\
		&\hspace{40pt}-\frac{\sigma_{0}^{2}}{P_{{\rm BS}}}\Tr(\hat{\mathbf{A}}_{k}\herm)\sum\nolimits _{i=1}^{K}\Tr(\bar{\mathbf{H}}\mathbf{F}_{i}\mathbf{F}_{i}\herm),\label{eq:EneqRk1}
	\end{align}
    where $\hat{\mathbf{B}}_{k}=\mathbf{\hat{\mathbf{X}}}_{k}\herm\mathbf{\hat{\mathbf{Y}}}_{k}^{-1}$ and we have used the equality $\Tr(\mathbf{U}\mathbf{V}) =\Tr(\mathbf{V}\mathbf{U})$.	We remark that $g_{k}(\mathbf{F})$ is a \emph{concave quadratic minorant} of $\tilde{R}_{k}(\mathbf{F})$. According to the SCA principle, the	precoder at iteration $n+1$, i.e., $\mathbf{F}^{(n+1)} \triangleq[\mathbf{F}_{1}^{(n+1)},\ldots,\mathbf{F}_{K}^{(n+1)}]$ is obtained as
	\begin{equation}
		\mathbf{F}^{(n+1)}=\underset{\mathbf{F}}{\argmax}\:\sum\nolimits_{k=1}^{K}g_{k}(\mathbf{F}).\label{eq:maxSCA}
	\end{equation}
	The quadratic nature of $g_{k}(\mathbf{F})$ leads to a closed-form solution described in the following theorem.
	\begin{thm}
		\label{thm:closed-form-beamformer}The closed-form solution for (\ref{eq:maxSCA}) is given by
		\begin{equation}
			\mathbf{F}=(\mu\mathbf{I}+\mathbf{\tilde{A}}\bar{\mathbf{H}})^{-1}\mathbf{\tilde{B}},\label{eq:cloded_form_F}\end{equation} where $\tilde{\mathbf{A}}=\diag(\omega_{1}\hat{\mathbf{A}}_{1}\herm,\cdots,\omega_{K}\hat{\mathbf{A}}_{K}\herm)$, $
		\tilde{\mathbf{B}}=\diag(\omega_{1}\hat{\mathbf{B}}_{1}\herm,\cdots,\omega_{K}\hat{\mathbf{B}}_{K}\herm)$,
		and $ \mu =\frac{\sigma_{0}^{2}}{P_{{\rm BS}}}\sum_{k=1}^{K}\omega_{k}\Tr(\hat{\mathbf{A}}_{k}\herm)$.
	\end{thm}
	\begin{IEEEproof}
		See \appref{proof-precoder-closed-form}.
	\end{IEEEproof}
	Finally, the proposed algorithm for optimizing the transmit precoder $\mathbf{W}$ by solving the equivalent problem $(\mathcal{P}_{2})$ is summarized in \algref{SCA}. The convergence of \algref{SCA} is stated in the following theorem.
	\begin{thm}
		\label{thm:SCA:convergence}Let $\{\mathbf{F}^{(n)}\}$ be the sequence produced by \algref{SCA}. Then the sequence $\{\sum_{k=1}^{K}\tilde{R}_{k}(\mathbf{F})\}$ converges. Further, any accumulation point of $\{\mathbf{F}^{(n)}\}$ is a stationary solution to $(\mathcal{P}_{2})$.
	\end{thm}
	\begin{IEEEproof}
		See \appref{SCA:convergence:proof}.
	\end{IEEEproof}
 \begin{algorithm}[!t]
		\SetAlgoLined
		\DontPrintSemicolon
		\SetKwRepeat{Do}{do}{while}
		\SetKwInput{Initialize}{Initialize}
		\SetKwInOut{Input}{Input}
		\SetKwInOut{Output}{Output}
        \Input{$\ensuremath{\mathbf{F}^{(0)}}$, $n\leftarrow0$.}
	\Repeat{$\ensuremath{convergence}$}
            {
			Compute $\ensuremath{\mathbf{F}^{(n+1)}}$ according to (\ref{eq:cloded_form_F})\;
			$n\leftarrow n+1$\;
		}
		\Output {$\mathbf{W}=\sqrt{\xi}\mathbf{H}\herm\mathbf{F}^{(n)}$, where $\xi=\frac{{P_{BS}}}{||\mathbf{H}\herm\mathbf{F}^{(n)}||^{2}}$.}
		\caption{SCA-based Method for Solving $(\mathcal{P}_{2})$\label{alg:SCA}}
	\end{algorithm}
    \subsection{RIS Optimization: $\boldsymbol{\theta}$-update}
    Given the demonstrated efficiency of optimizing $\mathbf{W}$ through $(\mathcal{P}_{2})$, it might seem intuitive to apply a similar idea for optimizing $\boldsymbol{\theta}$. Interestingly, however, this approach leads to poor performance, as will be shown later in the simulation results. Therefore, for optimizing $\boldsymbol{\theta}$, we use the original optimization problem in $(\mathcal{P}_{1})$.
	
	To proceed, if $\mathbf{W}$ is held fixed (i.e., $\mathbf{W}\leftarrow\mathbf{W}^{(\ell+1)}$), $(\mathcal{P}_{1})$ becomes
	\begin{equation}
		(\mathcal{P}_{3}):\Bigl\{\begin{array}{rl}
			\underset{\boldsymbol{\theta}}{\max} & R(\boldsymbol{\theta})\triangleq\sum_{k=1}^{K}\omega_{k}R_{k}(\boldsymbol{\theta})\\
			\st & \boldsymbol{\theta}\in\mathcal{Q},
		\end{array}
	\end{equation}
	where $\mathcal{Q}=\{\boldsymbol{\theta}\Bigl||\theta_{n}|=1,n=1,\cdots,N_{s}\}$ and 
	\begin{align}
        R_{k}(\boldsymbol{\theta}) &=\log\det(\sigma_{0}^{2}\mathbf{I}+\mathbf{H}_{k}\bar{\mathbf{W}}\mathbf{H}_{k}\herm)\nonumber \\
        &\quad-\log\det(\sigma_{0}^{2}\mathbf{I}+\mathbf{H}_{k}\bar{\mathbf{W}}_{k}\mathbf{H}_{k}\herm),\label{eq:rate-theta-givenW}
	\end{align}
    where $\bar{\mathbf{W}}=\sum_{j=1}^{K}\mathbf{W}_{j}\mathbf{W}_{j}\herm$ and $\bar{\mathbf{W}}_{k}= \bar{\mathbf{W}}-\mathbf{W}_{k}\mathbf{W}_{k}\herm=\sum_{j\neq k}^{K}\mathbf{W}_{j}\mathbf{W}_{j}\herm$. Note that we have dropped the dependency of the achievable rate of user $k$ on $\mathbf{W}$ to lighten the notation.
    \begin{rem}
    To the best of our knowledge, there are only a few solutions previously proposed for $(\mathcal{P}_{3})$, e.g. \cite{pan2020multicell}. In contrast, the MISO counterpart has seen several proposed approaches, mainly falling into two categories: gradient-based and SCA-based strategies (see e.g. \cite{guo2020weighted,zhou2020intelligent,farooq2022achievable} and references therein). In the former, projected gradient steps are performed to update $\boldsymbol{\theta}$. In the latter, a lower bound of $R_{k}(\boldsymbol{\theta})$  is achieved, based on which an improved solution of $\boldsymbol{\theta}$ is found. The main drawback of the SCA methods for solving $(\mathcal{P}_{3})$ lies in the difficulty of finding a tight lower bound and involves manipulations over large-size matrices. To resolve these shortcomings, we adopt a gradient-based approach for optimizing $\boldsymbol{\theta}$.
    \end{rem}
	Rather than aiming to find a stationary solution to $(\mathcal{P}_{3})$, which requires many projected gradient steps, we only execute a single projected gradient to update $\boldsymbol{\theta}$. In this regard, we recall that the \emph{conventional} projected gradient step is given by
    \begin{align}
        \boldsymbol{\theta}^{(\ell+1)} & =\Pi_{\mathcal{Q}}\bigl(\boldsymbol{\theta}^{(\ell)}+\alpha_{\ell+1}\nabla_{\boldsymbol{\theta}}R(\boldsymbol{\theta}^{(\ell)})\bigr),\label{eq:PG}
    \end{align}
    where $\Pi_{\mathcal{Q}}\left(.\right)$ denotes the projection of the argument onto set $\mathcal{Q}$, $\nabla_{\boldsymbol{\theta}}R(\boldsymbol{\theta}^{(\ell)})$ denotes the complex-valued gradient of $(\mathcal{P}_{3})$ at $\boldsymbol{\theta}^{(\ell)}$, and $\alpha_{\ell+1}>0$ is the step size. Through extensive numerical experiments, we have observed that this conventional projected gradient step tends to suffer significantly slow convergence rates for the proposed AO-based method. The main reason is the relatively small magnitude of $\nabla_{\boldsymbol{\theta}} R(\boldsymbol{\theta}^{(\ell)})$, which heavily depends on the indirect channel between BS and users via the RIS. Practically, this indirect channel is often weak, thereby hindering the convergence rate.
	
    To address this issue, we propose a scaled projected gradient (SPG) step, which shall be numerically shown to significantly accelerate the convergence speed. Specifically, our proposed SPG step is given by
	\begin{align}
		\boldsymbol{\theta}^{(\ell+1)} & =\Pi_{\mathcal{Q}}\bigl(\boldsymbol{\theta}^{(\ell)}+\alpha_{\ell+1}\boldsymbol{\boldsymbol{\Xi}}\nabla_{\boldsymbol{\theta}}R(\boldsymbol{\theta}^{(\ell)})\bigr),\label{eq:mPG}
	\end{align}
where $\boldsymbol{\boldsymbol{\Xi}}\succeq0$ accounts for the scaling of $\nabla_{\boldsymbol{\theta}}R(\boldsymbol{\theta}^{(\ell)})$. Based on our experience, we choose $\boldsymbol{\Xi}$ as follows
\begin{equation}
\boldsymbol{\boldsymbol{\Xi}}=\diag\bigl(1/|\nabla_{\boldsymbol{\theta}}R(\boldsymbol{\theta}^{(\ell)})|\bigr),\label{eq:SF}
\end{equation} 
where the division is element-wise. This essentially normalizes the complex-valued gradient by its modulus, resulting in an efficient way to find a proper step size, which shall be described shortly. The closed-form expression for $\nabla_{\boldsymbol{\theta}}R(\boldsymbol{\theta})$ is provided by the following theorem.
	\begin{thm}
		\label{thm:grad:theta}A closed-form expression for $\nabla_{\boldsymbol{\theta}}R(\boldsymbol{\theta})$ is given by (\ref{eq:dRwtheta}) shown at the top of the next page.
	\end{thm}
	\begin{IEEEproof}
		See \appref{proof-grad-theta-P1}.
	\end{IEEEproof}
	\begin{figure*}[t]
		\begin{multline}\nabla_{\boldsymbol{\theta}}R(\boldsymbol{\theta})=\sum\nolimits_{k=1}^{K}\omega_{k}\bigl(\vecd(\mathbf{U}_{k}\herm(\mathbf{H}_{k}\bar{\mathbf{W}}\mathbf{H}_{k}\herm+\sigma_{0}^{2}\mathbf{I})^{-1}\mathbf{H}_{k}\bar{\mathbf{W}}\mathbf{G}\herm)-\vecd(\mathbf{U}_{k}\herm(\mathbf{H}_{k}\bar{\mathbf{W}}_{k}\mathbf{H}_{k}\herm+\sigma_{0}^{2}\mathbf{I})^{-1}\mathbf{H}_{k}\bar{\mathbf{W}}_{k}\mathbf{G}\herm)\bigr)\label{eq:dRwtheta}
		\end{multline}
		\hrulefill{}
	\end{figure*}
	To complete the projected gradient step, we also need to find a step size. In related works, the line search procedure based on the Armijo--Goldstein condition is commonly used, which is given by \cite{boyd2004convex}
	\begin{gather}
		R(\boldsymbol{\theta}^{(\ell+1)})\geq R(\boldsymbol{\theta}^{(\ell)})+2\mathfrak{R}\{\nabla_{\boldsymbol{\theta}}R(\boldsymbol{\theta}^{(\ell)})\herm(\boldsymbol{\theta}^{(\ell+1)}-\boldsymbol{\theta}^{(\ell)})\}\nonumber \\
		+\frac{\alpha_{\ell+1}}{2}\Vert\boldsymbol{\theta}^{(\ell+1)}-\boldsymbol{\theta}^{(\ell)}\Vert^{2}.\label{eq:line-search-Amijo}
	\end{gather}
    However, we have observed in our empirical investigations that this method is inefficient for the considered problem as it tends to take several steps to find a proper step size, eventually leading to increased execution time. To address this issue, we propose a new line search as follows. Starting an initial value $\alpha_{\ell+1}:=c>0$ we keep decreasing $\alpha_{\ell+1}$ as $\alpha_{\ell+1}\leftarrow\eta\alpha_{\ell+1}$ where $\eta<1$ until the following inequality is met
	\begin{align}
		R(\boldsymbol{\theta}^{(\ell+1)}) & \geq R(\boldsymbol{\theta}^{(\ell)})+\frac{\beta}{2N_{s}}\parallel\boldsymbol{\theta}^{(\ell+1)}-\boldsymbol{\theta}^{(\ell)}\parallel^{2},\label{eq:proposed:LS}
	\end{align}
	where $\mathbf{\beta}$ is sufficiently small.
	\begin{lem}
		\label{lem:linesearch} The line search using the condition (\ref{eq:proposed:LS}) with the projected step in (\ref{eq:mPG}) is guaranteed to terminate after a finite number of steps from any positive initial step size.
	\end{lem}
	\begin{IEEEproof}
		See \appref{proof-LS}.
	\end{IEEEproof}
 
 In summary, \algref{AO} outlines the proposed AO-based algorithm for optimizing $\boldsymbol{\theta}$ and $\mathbf{W}$. Regarding the proposed line search above, the following two remarks are of note.
 \begin{algorithm}[!t]
		\SetAlgoLined
		\DontPrintSemicolon
		\SetKwRepeat{Do}{do}{while}
		\SetKwInput{Initialize}{Initialize}
		\SetKwInOut{Input}{Input}
		\SetKwInOut{Output}{Output}
        \Initialize{$\boldsymbol{\theta}^{(0)}$, $\eta<1$, $\epsilon>0$, $\beta>0$, $\ell \leftarrow {0}$.}	
		\Repeat{$R\bigl(\mathbf{W}^{(\ell+1)},\boldsymbol{\theta}^{(\ell+1)}\bigr)-R\bigl(\mathbf{W}^{(\ell)},\boldsymbol{\theta}^{(\ell)}\bigr)\leq\epsilon$
		}
		{
			Obtain $\mathbf{W}^{(\ell+1)}$ using \algref{SCA}\;
			Compute $\nabla_{\boldsymbol{\theta}}R\bigl(\boldsymbol{\theta}^{(\ell)}\bigr)$ according to (\ref{eq:dRwtheta})\;
			Compute $\boldsymbol{\Xi}$ according to (\ref{eq:SF})\;
			$\alpha_{\ell+1}=\frac{1}{R(\boldsymbol{\theta}^{(\ell)})}$ \tcp*{ initial step size; $R\bigl(\boldsymbol{\theta}^{(\ell)}\bigr)\equiv R\bigl(\boldsymbol{\theta}^{(\ell)},\mathbf{W}^{(\ell+1)}\bigr)$}
			\Repeat{$R\bigl(\boldsymbol{\theta}^{(\ell+1)}\bigr)\geq R\bigl(\boldsymbol{\theta}^{(\ell)}\bigr)+\frac{\beta}{2N_{s}}\parallel\boldsymbol{\theta}^{(\ell+1)}-\boldsymbol{\theta}^{(\ell)}\parallel^{2}$}
			{
				$\boldsymbol{\theta}^{(\ell+1)}=\Pi_{\mathcal{Q}}\bigl(\boldsymbol{\theta}^{(\ell)}+\alpha_{\ell+1}\boldsymbol{\boldsymbol{\Xi}}\nabla_{\boldsymbol{\theta}}R(\boldsymbol{\theta}^{(\ell)})\bigr)$
				\tcp*{ SPG step }
				$\alpha_{\ell+1}\leftarrow\eta\alpha_{\ell+1}$ \tcp*{ reduce step size}
			}
			$\ell\leftarrow \ell+1$\;
		}
		\caption{AO-based Algorithm for Solving $(\mathcal{P}_{1})$.}\label{alg:AO}
	\end{algorithm}
\begin{rem}
        Our motivation to adopt the above line-search approach, as opposed to (\ref{eq:line-search-Amijo}), is as follows. First, by noting that $\parallel\boldsymbol{\theta}^{(\ell+1)}-\boldsymbol{\theta}^{(\ell)}\parallel^{2}\leq\parallel\boldsymbol{\theta}^{(\ell+1)}\parallel^{2}+\parallel\boldsymbol{\theta}^{(\ell)}\parallel^{2}=2N_{s}$. Thus, by assigning $\beta$ to a small value, the line search routine often takes fewer steps, which could lead to a moderate objective increase. This is contrary to the traditional projected gradient step, which strikes to maximally increase the objective, which is normally the case when using (\ref{eq:line-search-Amijo}). In other words, we slightly improve the objective by the $\boldsymbol{\theta}$-update and let the $\mathbf{W}$-update drive the WSR improvement of the overall objective. This is because digital transmit beamforming can highly coordinate inter-user interference more than passive beamforming.
\end{rem}
	\begin{rem}
        The choice of the initial value plays a crucial role in determining the number of iterations required to find a proper value for the step size $\alpha_{\ell+1}$. A popular way is to carry forward the step size of the previous iteration as a starting point for the line search procedure in the next iteration. However, our extensive numerical experiments reveal that this method is inefficient since, after a few AO iterations, the step size becomes vanishingly small, and thus, the AO progresses very slowly. To overcome this issue, we propose initializing the step size as $\alpha_{\ell+1}=\nicefrac{1}{R(\boldsymbol{\theta}^{(\ell)})}$ for the $\ell$-th iteration of the proposed AO-based algorithm. Given that $R(\boldsymbol{\theta}^{(\ell)})$ increases after each iteration of the AO algorithm, the initial value for $\alpha_{\ell+1}$ decreases accordingly. This initialization approach has been observed, in our numerical experiments, to often finish the line search using (\ref{eq:proposed:LS}) within a single iteration, thereby significantly enhancing the convergence speed.
	\end{rem}
	
	\subsection{On the Numerical Efficiency of $\boldsymbol{\theta}$-update using $(\mathcal{P}_{2})$\label{subsec:On-the-Numerical_P2}}	
	In the preceding subsection, we have stated that for the $\boldsymbol{\theta}$-update step, the problem formulation given in $(\mathcal{P}_{1})$ \emph{is preferred over that} in $(\mathcal{P}_{2})$. This choice is guided by an empirical analysis of the Lipschitz constant of the gradients of the WSR w.r.t. $\boldsymbol{\theta}$ for both problem formulations. To understand this, let us fix $\mathbf{F}$ and rewrite  (\ref{eq:rate-equivalent}) as a function of $\boldsymbol{\theta}$, i.e.
	\begin{multline}
		\tilde{R}_{k}(\boldsymbol{\theta})=\log\det\biggl(\mathbf{I}+\bar{\mathbf{H}}_{k}(\boldsymbol{\theta})\mathbf{F}_{k}\mathbf{F}_{k}\herm\bar{\mathbf{H}}_{k}(\boldsymbol{\theta})\herm\\
		\,\times\Bigl(\sum\nolimits _{j\neq k}^{K}\bar{\mathbf{H}}_{k}(\boldsymbol{\theta})\mathbf{F}_{j}\mathbf{F}_{j}\herm\bar{\mathbf{H}}_{k}(\boldsymbol{\theta})\herm\\
       \,+\frac{\sigma_{0}^{2}}{P_{{\rm BS}}}\sum\nolimits _{i=1}^{K}\Tr(\mathbf{F}_{i}\herm\bar{\mathbf{H}}(\boldsymbol{\theta})\mathbf{F}_{i})\mathbf{I}\Bigr)^{-1}\biggr),\label{eq:rate-equivalent-theta}
	\end{multline}
where the dependency of the equivalent channels on $\boldsymbol{\theta}$ is explicitly shown.. For the sake of brevity, we restrict ourselves to a RIS-aided MISO system and derive the gradient of $\tilde{R}(\boldsymbol{\theta})$ in \appref{rate-equivalent-grad}. To gain insights into the performance of gradient-based methods when applied to $\tilde{R}_{k}(\boldsymbol{\theta})$, it is crucial to analyze the Lipschitz constant of its gradient. In this regard, we recall that a differentiable function $f$ is said to have an $L$-Lipschitz continuous gradient for parameter $L>0$ if $||\nabla f(\mathbf{x})-\nabla f(\mathbf{y})||\leq L||\mathbf{x}-\mathbf{y}||$ for any two points $\mathbf{x}$ and $\mathbf{y}$ within its domain. In general, the number of iterations required to converge tends to increase with the Lipschitz constant of the gradient \cite{jones1993lipschitzian}. Although closed-form expressions for $R(\boldsymbol{\theta})$ and $\tilde{R}(\boldsymbol{\theta})$ are given in (\ref{eq:dRwtheta}) and (\ref{eq:rate-equivalent-grad}), respectively, finding the Lipschitz constant is still very challenging. Instead, we randomly generated $10^{6}$ samples of $\boldsymbol{\theta}$ over the feasible set and empirically estimate the gradient's Lipschitz constant accordingly, which is plotted in \figref{FigLptz}.
      \begin{figure}[!t]
     \centering{}\includegraphics[scale=0.75]{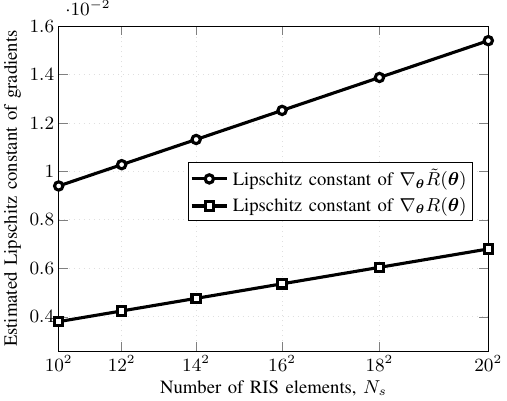}\caption{Estimated Lipschitz constants of $\nabla_{\boldsymbol{\theta}}\tilde{R}(\boldsymbol{\theta})$ and $\nabla_{\boldsymbol{\theta}}R(\boldsymbol{\theta})$ for a RIS-aided MISO system with $N_{t}=64$, $K=4$, and $P_{{\rm BS}}=30~{\rm dBm}$.}\label{fig:FigLptz}
	\end{figure}    
    As can be seen clearly, the Lipschitz constant of $\nabla_{\boldsymbol{\theta}}\tilde{R}(\boldsymbol{\theta})$ is approximately two times larger than that of $\nabla_{\boldsymbol{\theta}}R(\boldsymbol{\theta})$. Thus, when gradient-based optimization methods are employed within the context of $(\mathcal{P}_{2})$, they tend to require a greater number of iterations to achieve convergence compared to their application in $(\mathcal{P}_{1})$. This can be attributed to the inherent properties of the gradient's Lipschitz continuity; a higher Lipschitz constant implies that the gradient exhibits more significant changes over small variations in $\boldsymbol{\theta}$, thereby impacting the convergence rate. \figref{FigLptz} also shows that the Lipschitz constant of $\nabla_{\boldsymbol{\theta}}R(\boldsymbol{\theta})$ is relatively small, which justifies our SPGM proposed in the preceding subsection.
 \subsection{Convergence Analysis}
	The convergence of \algref{AO} is as follows. According to \thmref{SCA:convergence}, an accumulation point of \algref{SCA} is a stationary solution to $(\mathcal{P}_{2})$. Thus $\mathbf{W}^{(\ell)}$ is a nontrival stationary solution to $(\mathcal{P}_{1})$ for a given $\boldsymbol{\theta}$, as shown by \cite[Theorem 2]{zhao2023rethinking}. It is obvious that \algref{AO} generates a nondecreasing sequence $\{R(\mathbf{W}^{(\ell+1)},\boldsymbol{\theta}^{(\ell+1)})\}$, whose convergence is guaranteed due to its continuity over the closed feasible set. Following similar arguments in \cite[Chap. 2.7]{Bertsekas1999NP}, it can be shown that any accumulation point of the sequence $\{\mathbf{W}^{(\ell)},\boldsymbol{\theta}^{(\ell)}\}$ is a stationary solution to $(\mathcal{P}_{1})$.
	
    \subsection{Complexity Analysis\label{subsec:Complexity-Analysis}}
    We now analyze the computational complexity of \algref{AO} by counting the number of required complex multiplications \cite{Raphel}. Our analysis focuses on operations that contribute most significantly to the overall computational complexity, taking into the fact that typically in practice $N_{s}\gg N_{t}\gg KN_{r}\geq KN_{d}\geq K$.
    
    The dominant complexity of our proposed method depends heavily on lines 3 and 4 of \algref{AO}. First, we need to generate matrices $\mathbf{H}$ and $\bar{\mathbf{H}}$ whose computational complexities are $\mathcal{O}(N_{s}N_{t}N_{r}K)$ and $\mathcal{O}(N_{t}N_{r}^{2}K^{2})$, respectively. The complexity of $\mathbf{W}$-update via \algref{SCA} is dominated by (\ref{eq:cloded_form_F}), which is of $\mathcal{O}(N_{r}^{3}K^{3})$ multiplications.
    
    Next, we discuss efficient steps to compute $\nabla_{\boldsymbol{\theta}}R\left(\boldsymbol{\theta}\right)$ in line 4 of \algref{AO}. Let us first define $\mathbf{Z}_{k}= \mathbf{H}_{k}\bar{\mathbf{W}}\mathbf{H}_{k}\herm +\sigma_{0}^{2}\mathbf{I}$, $\tilde{\mathbf{Z}}_{k} =\mathbf{H}_{k}\bar{\mathbf{W}}_{k}\mathbf{H}_{k}\herm +\sigma_{0}^{2}\mathbf{I}$, $\mathbf{J}_{k}=\mathbf{H}_{k}\bar{\mathbf{W}}$, and $\tilde{\mathbf{J}}_{k} =\mathbf{H}_{k}\bar{\mathbf{W}_{k}}$. The initial step involves computing and storing the matrices $\mathbf{E}_{j}=\mathbf{H}_{k}\mathbf{W}_{j}$, $\mathbf{M}_{j}=\mathbf{E}_{j}\mathbf{E}_{j}\herm$, and $\mathbf{N}_{j}=\mathbf{E}_{j}\mathbf{W}_{j}\herm$, which require complexities of $\mathcal{O}(N_{t}N_{r}N_{d}K)$, $\mathcal{O}(N_{r}^{2}N_{d}K)$, and $\mathcal{O}(N_{t}N_{r}N_{d}K)$, respectively. Then we define $\mathbf{Z}_{k}\equiv\sum_{j=1}^{K}\mathbf{M}_{j}+\sigma_{0}^{2}\mathbf{I}$, $\tilde{\mathbf{Z}}_{k}\equiv\sum_{j\neq k}^{K}\mathbf{M}_{j}+\sigma_{0}^{2}\mathbf{I}$, $\mathbf{J}_{k}\equiv\sum_{j=1}^{K}\mathbf{N}_{j}$, and $\tilde{\mathbf{J}}_{k}\equiv\sum_{j\neq k}^{K}\mathbf{N}_{j}$. Second, we compute $\mathbf{Z}_{k}^{-1}$ and $\tilde{\mathbf{Z}}_{k}^{-1}$, each requiring $\mathcal{O}(N_{r}^{3})$ multiplications. The next step involves calculating $\mathbf{Z}_{k}^{-1}\mathbf{J}_{k}$ and $\tilde{\mathbf{Z}}_{k}^{-1}\tilde{\mathbf{J}}_{k}$, with each calculation requiring $\mathcal{O}(N_{t}N_{r}^{2})$ multiplications. Finally, each of the terms $\vecd(\mathbf{U}_{k}\herm(\mathbf{Z}_{k}^{-1}\mathbf{J}_{k})\mathbf{G}\herm)$ and $\vecd(\mathbf{U}_{k}\herm(\tilde{\mathbf{Z}}_{k}^{-1}\tilde{\mathbf{J}}_{k})\mathbf{G}\herm)$ needs $\mathcal{O}(N_{s}N_{t}(N_{r}+1))$ multiplications. Given that these calculations are performed across all $K$ user channels, the total complexity for computing $\nabla_{\boldsymbol{\theta}}R(\boldsymbol{\theta})$ is thus $\mathcal{O}(N_{t}N_{r}N_{d}K^{2}+N_{r}^{2}N_{d}K^{2}+2N_{r}^{3}K+N_{t}N_{r}^{2}K+2N_{s}N_{t}K(N_{r}+1))\approx \mathcal{O}(N_{t}N_{r}N_{d}K^{2}+2N_{s}N_{t}N_{r}K)$. Meanwhile, the complexity of the line search (i.e., lines $7-10$ in \algref{AO}) is dominated by computing $\mathbf{H}$, which is $\mathcal{O}(I_{\theta}(N_{s}N_{t}N_{r}K))$, where $I_{\theta}$ is the average number of iterations employed in the line-search. Therefore, the total dominant complexity of one outer iteration (i.e., lines $2-12$ in \algref{AO}) is given by 
	\begin{align}
		\mathcal{O}\bigl(N_{t}N_{r}N_{d}K^{2}+I_{\theta}(N_{s}N_{t}N_{r}K)+I_{w}(N_{r}^{3}K^{3})\bigr),\label{eq:C_MIMO}
	\end{align} where $I_{w}$ denotes the number of inner iterations (i.e., lines $2-5$) in \algref{SCA}. The above expression clearly indicates that the complexity of computing $\nabla_{\boldsymbol{\theta}}R(\boldsymbol{\theta})$ \emph{grows linearly} with $N_{s}$, i.e., $O(N_{s})$. This contradicts the commonly held belief within the existing literature that the complexity is to be quadratic, i.e., $O(N_{s}^{2})$. For further	details, we refer the interested readers to the complexity analyses in \cite{guo2020weighted, JC2021Chen, yu2019miso}.
 
    \section{Simulation Results\label{sec:SimResults}}
    \subsection{Simulation Settings}
    In simulation settings, we consider uniform linear arrays (ULAs) for RIS-aided MIMO systems, where the BS, RIS, and users are equipped with half-wavelength inter-element spacing. Unless otherwise stated, we assume that the BS is equipped with $N_{t}=64$ antennas and serves $K=4$ users, each equipped with $N_{r}=2$ antennas to receive $N_{d}=2$ data stream via a RIS consisting of $N_{s}=400$ reflecting elements. Additionally, the BS has the maximum power of $P_{{\rm BS}}=30~{\rm dBm}$ and the noise power is set as $\sigma_{0}^{2}=-90~{\rm dBm}$. The BS and RIS are located at the coordinates $(0,0)$ and $(200~{\rm m},0)$, respectively, while the users are randomly distributed within a $10~{\rm m}$ radius circle centered at $(200~{\rm m},30~{\rm m})$ from the BS. The users' priority weights are randomly generated and fixed for all numerical experiments as follows: $\omega_{1}=0.2449$, $\omega_{2}=0.2509$, $\omega_{3}=0.2570$, and $\omega_{4}=0.2472$. Furthermore, we consider the Rayleigh fading model for the direct link BS-UEs channel, while the Rician fading model is considered for the BS-RIS and RIS-UEs channels. We follow the 3GPP guideline for modeling the large-scale fading propagation, where the path-loss formulas for line-of-sight (LoS) and non-LoS (NLoS) are given by 
	\[
	{P_{loss}}(d)[{\rm dBm}]=\begin{cases}
		35.6+22.0\log_{10}(d), & {\rm LoS}\\
		32.6+36.7\log_{10}(d), & {\rm nLoS}
	\end{cases}
	\]	
	where $d$ is the distance in meters. The channels $\mathbf{G}$, $\mathbf{U}_{k}$, and $\mathbf{D}_{k}$ are modeled as follows:
	\begin{align*}
    \mathbf{G}& =\sqrt{\mathcal{L}_{1}\kappa_{1}}\mathbf{a}_{N_{s}}\left(\vartheta\right)\mathbf{a}_{N_{t}}\left(\varphi\right)^{H}+\sqrt{\mathcal{L}_{1}\kappa_{2}}\mathbf{\bar{G}},\\
    \mathbf{U}_{k}& =\sqrt{\mathcal{L}_{2,k}\kappa_{1}}\mathbf{a}_{N_{r}}\left(\psi_{k}\right)\mathbf{a}_{N_{s}}\left(\phi_{k}\right)^{H}+\sqrt{\mathcal{L}_{2,k}\kappa_{2}}\mathbf{\bar{U}}_{k},\\
    \mathbf{D}_{k}&=\mathcal{L}_{3,k}\mathbf{\bar{D}}_{k},
	\end{align*}
    where $\mathcal{L}_{1}$, $\mathcal{L}_{2,k}$, and $\mathcal{L}_{3,k}$ denote the path-losses corresponding to the links from the BS to RIS, the RIS to user $k$, and the BS to user $k$, respectively, where $\mathcal{L}_{(\cdot)}=10^{(P_{loss}-30)/10}$; $\mathbf{\bar{G}}$, $\mathbf{\bar{U}}_{k}$, and $\mathbf{\bar{D}}_{k}$ are the corresponding NLoS channels modeled as Rayleigh fading whose entries are chosen from $\mathcal{CN}(0,1)$; $\kappa_{1}=1-\kappa_{2}$ and $\kappa_{2}=1/(1+\kappa)$, where $\kappa$ is the Rician factor and we set $\kappa=10$; $\mathbf{a}_{N_{s}}(\vartheta)$, $\mathbf{a}_{N_{t}}(\varphi)$, $\mathbf{a}_{N_{s}}(\phi_{k})$, and $\mathbf{a}_{N_{r}}(\psi_{k})$ are the antenna steering vectors with $\vartheta$, $\varphi$, $\phi_{k}$, and $\psi_{k}$ as the corresponding angular parameters. Furthermore, we employ the same initial random vector $\boldsymbol{\theta}$ for all algorithms. We set $\eta=0.5$, $\epsilon=10^{-5}$, $\beta=10^{-7}$, and average all simulation outcomes over $100$ independent channel realizations.
    \subsection{Baseline Schemes}
    To assess the effectiveness of our proposed algorithm in this paper, we compare it against the following baseline schemes (BLSs):
    \begin{enumerate}
        \item BLS1: This scheme is a slightly modified version of \algref{AO}, employing the conventional PG-based method with the Armijo-Goldstein line search technique in (\ref{eq:line-search-Amijo}) for the $\boldsymbol{\theta}$-update. We include this scheme to demonstrate the effectiveness of our proposed SPG step in \algref{AO}.
     \item BLS2: This scheme is another variant of \algref{AO}. However, the $\boldsymbol{\theta}$-update is performed using the equivalent problem $(\mathcal{P}_{2})$. We note that the gradient w.r.t $\boldsymbol{\theta}$ of $\tilde{R}(\boldsymbol{\theta})$ is provided in (\ref{eq:rate-equivalent-grad}). The purpose of including this scheme is to illustrate our novelty of using $(\mathcal{P}_{1})$ for RIS optimization rather than $(\mathcal{P}_{2})$ as discussed in \subsecref{On-the-Numerical_P2}.
    \item BCD algorithm \cite{guo2020weighted}: It serves as a comparative benchmark against our proposed scheme. Note that this scheme only applies to the special case of RIS-aided MISO systems.
    \item WMMSE-MM \cite{pan2020multicell}: This scheme deals with RIS-aided MIMO systems and is included for comparison to demonstrate the effectiveness of our proposed algorithm in handling systems with very large $N_{t}$ and $N_{s}$.
    \item Random Phase: In this scheme, $\boldsymbol{\theta}$ is randomly generated and remains fixed while $\mathbf{W}$ is optimized using \algref{SCA}.
    \item Without RIS: This baseline only relies on the direct link between the BS and UEs without RIS.
    \end{enumerate} 
   \begin{figure}[!t]
    \begin{centering}
    \subfloat[\label{subfig:ConvMIMO}]{\centering{}\includegraphics[scale=0.75]{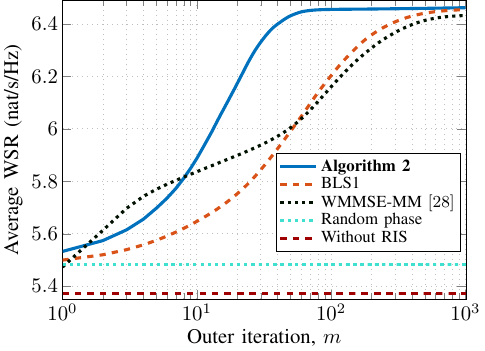}}
    \par\end{centering}\vspace{-8pt}
    \begin{centering}
    \subfloat[\label{subfig:ConvMISO}]{\centering{}\includegraphics[scale=0.75]{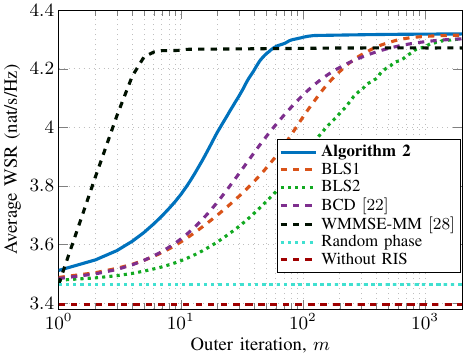}}
    \par\end{centering}
    \caption{Algorithms convergence behavior results for (a). RIS-aided MIMO system with $N_{s}=400$, $N_{t}=64$, $K=4$, $N_{r}=N_{d}=2$, and $P_{{\rm BS}}=30~{\rm dBm}$  (b). RIS-aided MISO system with $N_{s}=400$, $N_{t}=64$, $K=4$, and $P_{{\rm BS}}=30~{\rm dBm}$.}\label{fig:ConvAlgs}
    \end{figure}
     \begin{figure}[!t]
    \begin{centering}
    \subfloat[\label{subfig:RvNrisMIMO}]{\centering{}\includegraphics[scale=0.75]{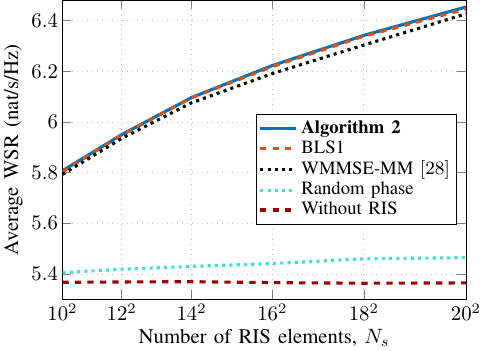}}%
    \par\end{centering}\vspace{-8pt}
    \begin{centering}
    \subfloat[\label{subfig:RvNrisMISO}]{\centering{}\includegraphics[scale=0.75]{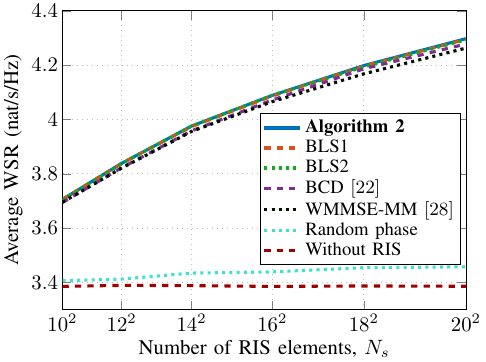}}%
    \par\end{centering}
    \caption{Achievable WSR versus $N_{s}$ for (a). RIS-aided MIMO system with $N_{t}=64$, $K=4$, $N_{r}=N_{d}=2$, and $P_{{\rm BS}}=30~{\rm dBm}$ (b). RIS-aided MISO system with $N_{t}=64$, $K=4$, and $P_{{\rm BS}}=30~{\rm dBm}$.}\label{fig:WSRvsNris}
    \end{figure} 
 \par In \figref{ConvAlgs}, we illustrate the convergence behavior of \algref{AO} and the other benchmark schemes. From \subfigref{ConvMIMO}, it is evident that our proposed algorithm, utilizing the SPGM for optimizing $\boldsymbol{\theta}$, achieves faster convergence compared to BLS1, which employs the conventional PG method. This significant difference underscores the efficiency of our proposed SPGM. In \subfigref{ConvMISO}, we observe that while the WMMSE-MM algorithm stabilizes relatively fast, \algref{AO} achieves better performance upon full convergence. However, it is important to note that, in both \subfigref{ConvMIMO} and \subfigref{ConvMISO}, the convergence comparison does not account for the per iteration complexity of each algorithm, which significantly affects the overall computational complexity and run-time. This issue is studied in great detail in Figs. \ref{fig:CPLXvsNris}, \ref{fig:CPLXvsTx}, and \ref{fig:TvsNris} in the sequel. Moreover, as shown in \subfigref{ConvMISO}, our proposed algorithm converges with the highest WSR compared to the benchmark algorithms. It is also worth noting that  BLS2, which optimizes $\boldsymbol{\theta}$ and $\mathbf{W}$ using the equivalent problem $(\mathcal{P}_{2})$, despite employing the SPGM, exhibits slower convergence compared to other algorithms. This observation confirms the analysis presented in \subsecref{On-the-Numerical_P2}. Importantly, since the overall computational complexity is heavily influenced by both the convergence speed and actual computational loads within the algorithm, the proposed algorithm eventually has the lowest complexity compared to the benchmark algorithms, as presented shortly.\par 
 \figref{WSRvsNris} shows the impact of $N_{s}$ on the achievable WSR of various schemes. As can be seen clearly,  algorithms employing optimized $\boldsymbol{\theta}$ exhibit significant performance improvements as $N_{s}$ increase. In contrast, the WSR of the Random Phase scheme, which does not optimize $\boldsymbol{\theta}$, barely increases with $N_{s}$. \figref{WSRvsNris} clearly indicates that the proposed algorithm slightly outperforms the benchmark schemes.	
 \begin{figure}[!t]
    \begin{centering}
    \subfloat[\label{subfig:CvNrisMIMO}]{\centering{}\includegraphics[scale=0.75]{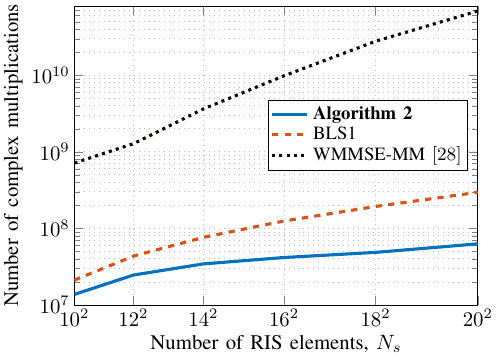}}%
    \par\end{centering}\vspace{-8pt}
    \begin{centering}
    \subfloat[\label{subfig:CvNrisMISO}]{\centering{}\includegraphics[scale=0.75]{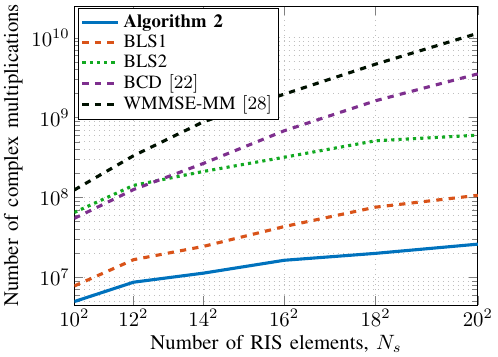}}%
    \par\end{centering}
    \caption{Number of complex multiplications versus $N_{s}$ for (a). RIS-aided MIMO system with $N_{t}=64$, $K=4$, $N_{r}=N_{d}=2$ and $P_{{\rm BS}}=30~{\rm dBm}$ (b). RIS-aided MISO system with $N_{t}=64$, $K=4$, and $P_{{\rm BS}}=30~{\rm dBm}$.}\label{fig:CPLXvsNris}
    \end{figure} 
 \par In \figref{CPLXvsNris}, we examine the impact of $N_{s}$ on the number of complex multiplications. It is noteworthy that the per-outer iteration complexity of BLS1 is the same as the expression provided in (\ref{eq:C_MIMO}). The only difference is in the number of iterations for convergence. We also remark that since BLS2 utilizes the gradient of the equivalent problem given in \appref{rate-equivalent-grad}, its per-outer iteration complexity is $\mathcal{O}(2N_{s}N_{t}K^{2}+I_{\theta}N_{s}N_{t}KI_{w}K^{3})$, whereas the per-outer iteration complexity of the BCD algorithm is $\mathcal{O}(N_{s}^{2}K^{2}+I_{\theta}(2N_{s}N_{t}K+N_{t}^{2}K))$ as provided in \cite{guo2020weighted}, while the complexity of WMMSE-MM \cite{pan2020multicell} is  $\mathcal{O}(N_{s}^{3}+N_{s}^{2}N_{t}+I_{\theta}N_{s}^{2}+I_{w}N_{t}^{3})$.  From the results presented in \subfigref{CvNrisMIMO} for MIMO systems and \subfigref{CvNrisMISO} for MISO systems, it is evident that the proposed algorithm significantly reduces the overall computational complexity compared to other baseline algorithms.  
   \begin{figure}[!t]
    \begin{centering}
    \subfloat[\label{subfig:CvsTxMIMO}]{\centering{}\includegraphics[scale=0.75]{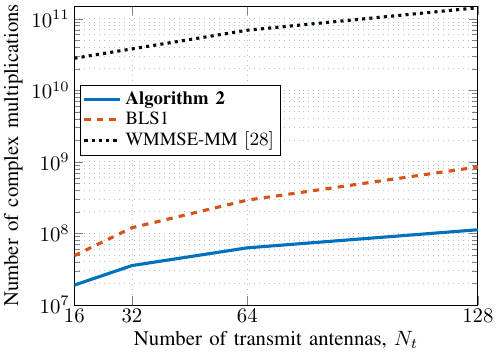}}%
    \par\end{centering}\vspace{-8pt}
    \begin{centering}
    \subfloat[\label{subfig:CvsTxMISO}]{\centering{}\includegraphics[scale=0.75]{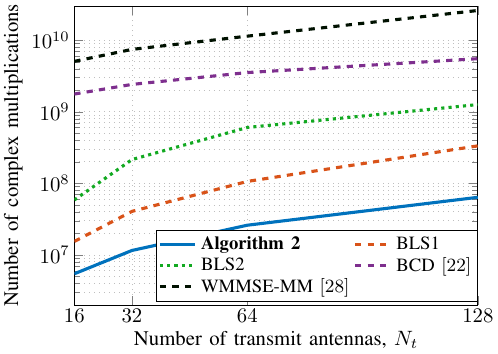}}%
    \par\end{centering}
    \caption{Number of complex multiplications versus $N_{t}$ for (a). RIS-aided MIMO system with $N_{s}=400$, $K=4$, $N_{r}=N_{d}=2$ and $P_{{\rm BS}}=30~{\rm dBm}$ (b). RIS-aided MISO system with $N_{s}=400$, $K=4$, and $P_{{\rm BS}}=30~{\rm dBm}$.}\label{fig:CPLXvsTx}
\end{figure} 
\begin{figure}[!t]
    \begin{centering}
    \subfloat[\label{subfig:TvsNrisMIMO}]{\centering{}\includegraphics[scale=0.75]{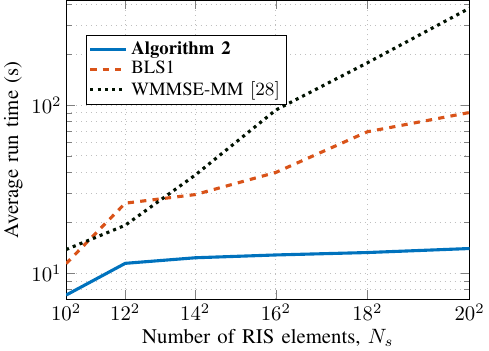}}%
    \par\end{centering}\vspace{-8pt}
    \begin{centering}
    \subfloat[\label{subfig:TvsNrisMISO}]{\centering{}\includegraphics[scale=0.75]{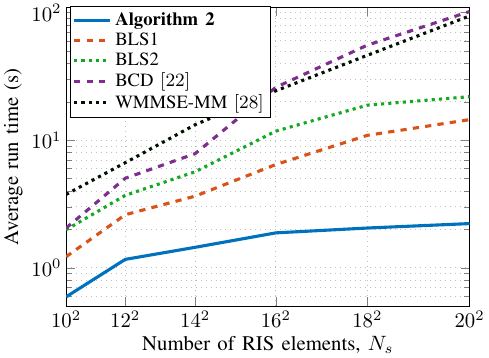}}%
    \par\end{centering}
    \caption{Average running time versus $N_{s}$ for (a). RIS-aided MIMO system with $N_{t}=64$, $K=4$, $N_{r}=N_{d}=2$, and $P_{{\rm BS}}=30~{\rm dBm}$ (b). RIS-aided MISO system with $N_{t}=64$, $K=4$, and $P_{{\rm BS}}=30~{\rm dBm}$.}\label{fig:TvsNris}
\end{figure}
 \par In \figref{CPLXvsTx}, we investigate how the algorithms in comparison scale with the number of transmit antennas, $N_{t}$, regarding the required number of complex multiplications to converge. For both MIMO and MISO setups, it can be seen that the proposed algorithm shows a linear increase with $N_{t}$, which confirms our complexity analysis presented above. Moreover, the proposed algorithm achieves a significantly lower complexity than the benchmark schemes in all considered cases. Specifically, as shown in \figref{CPLXvsTx}, the proposed algorithm achieves a significantly lower complexity than the BSL1 due to its faster convergence, requiring fewer per-outer iterations. Additionally, as shown in \figref{CPLXvsTx}, the WMMSE-MM algorithm exhibits the highest computational complexity, highlighting its ineffectiveness for large-scale RIS-aided MIMO systems.   

 \par \figref{TvsNris} illustrates the effect of $N_{s}$ on the average problem-solving time across different algorithms in comparison. Particularly, \subfigref{TvsNrisMIMO} demonstrates that the execution time for the proposed algorithm slightly grows with the increase of $N_{s}$, whereas for the BSL1, its run-time increases significantly with higher values of $N_{s}$. In \subfigref{TvsNrisMISO}, we also observe that \algref{AO} needs extremely less run time to converge compared to the WMMSE-MM algorithm, despite requiring more outer iterations as seen earlier in \subfigref{ConvMISO}. This is because each iteration of \algref{AO} involves lower-dimensional matrices, resulting in reduced computation time. Furthermore, in \subfigref{TvsNrisMISO}, we observe that the proposed algorithm exhibits the shortest run-time among all five compared algorithms. The collective results presented in \figref{CPLXvsNris}, \figref{CPLXvsTx}, and \figref{TvsNris} validate our complexity analysis, demonstrating that the complexity of the proposed algorithm scales linearly with  $N_{t}$ and  $N_{s}$. These results underscore the efficacy of our optimization techniques, particularly in handling large-scale RIS-assisted systems without incurring a large increase in overall execution time and computational complexity. 
  \section{Conclusion  and Future Work}\label{sec:Conclusion} 
 In this paper, we have presented an efficient algorithm for the joint optimization of the precoders at the BS and phase shifts at the RIS to maximize the WSR for large-scale RIS-assisted MU-MIMO downlink systems. Specifically, we utilized an equivalent lower-dimensional reformulation of the WSR maximization problem to derive a closed-form solution for transmit beamforming design based on the SCA approach. Furthermore, we proposed the SPGM algorithm to optimize the RIS phase shifts, incorporating a novel line search method to determine the step sizes for the SPGM. Moreover, we conducted a complexity analysis and showed that the complexity of the proposed algorithm scales linearly with the number of BS antennas and RIS elements. Numerical simulations demonstrated that the proposed algorithm outperforms baseline schemes in terms of WSR performance while attaining significant reductions in both time and computational complexity. For future work, CSI estimation methods are always worth studying. It is also interesting to evaluate our proposed algorithm in practical scenarios, such as those involving imperfect CSI and hardware impairments.
	
    \appendix{}	
    \subsection{Proof of (\ref{eq:enequlityXY})\label{app:proof-key-inequality}}	
	We now provide a more elegant proof of (\ref{eq:enequlityXY}). First, by the matrix inversion lemma \cite[Lemma 2.3]{hjorungnes2011complex} we have 
	\begin{equation}
		(\mathbf{I}+\mathbf{X}_{k}\herm\mathbf{Y}_{k}^{-1}\mathbf{X}_{k})^{-1}=\mathbf{I}-\mathbf{X}_{k}\herm(\mathbf{Y}_{k}+\mathbf{X}_{k}\mathbf{X}_{k}\herm)^{-1}\mathbf{X}_{k}.
	\end{equation}
	Let $\mathbf{P}_{k}=\mathbf{Y}_{k}+\mathbf{X}_{k}\mathbf{X}_{k}\herm$ and define $f(\mathbf{X}_{k},\mathbf{P}_{k})=\log\det(\mathbf{I}-\mathbf{X}_{k}\herm\mathbf{P}_{k}^{-1}\mathbf{X}_{k})$. Then we show that $f(\mathbf{X}_{k},\mathbf{P}_{k})$ is \emph{jointly concave} with $\mathbf{X}_{k}$ and $\mathbf{P}_{k}$, which is equivalent to showing that the following constraint 
	\begin{equation}
		f(\mathbf{X}_{k},\mathbf{P}_{k})\geq t
	\end{equation}
	is convex. It is easy to see that the above constraint can be equivalently expressed as 
	\begin{subequations}
		\begin{align}
			\log\det(\mathbf{I}-\mathbf{Q}_{k}) & \geq t\\
			\mathbf{I}-\mathbf{Z}_{k} & \succeq0\\
			\mathbf{Q}_{k}-\mathbf{X}_{k}\herm\mathbf{P}_{k}^{-1}\mathbf{X}_{k} & \succeq0.\label{eq:schurform}
		\end{align}
	\end{subequations}
	By the Schur complement, (\ref{eq:schurform}) is equivalent to the following linear matrix inequality
	\begin{equation}
		\begin{bmatrix}\mathbf{Q}_{k} & \mathbf{X}_{k}\herm\\
			\mathbf{X}_{k} & \mathbf{P}_{k}
		\end{bmatrix}\succeq0.\label{eq:LMI:schur}
	\end{equation}
	It is now clear that the concavity of $f(\mathbf{X}_{k},\mathbf{P}_{k})$ is due to (\ref{eq:LMI:schur}) and the concavity of $\log\det(\mathbf{I}-\mathbf{Q}_{k})$ over the domain $\mathbf{I}-\mathbf{Q}_{k}\succeq0$. Thus, the affine approximation of $f(\mathbf{X}_{k},\mathbf{P}_{k})$ around $(\hat{\mathbf{X}}_{k},\hat{\mathbf{P}}_{k}=\hat{\mathbf{Y}}_{k}+\hat{\mathbf{X}}_{k}\hat{\mathbf{X}}_{k}\herm$) produces an upper bound, i.e. 
	\begin{align}
		f(\mathbf{X}_{k},\mathbf{P}_{k})&\leq f(\hat{\mathbf{X}}_{k},\hat{\mathbf{P}}_{k})+\Tr\bigl(\nabla_{\mathbf{P}_{k}}f(\hat{\mathbf{X}}_{k},\hat{\mathbf{P}}_{k})(\mathbf{P}_{k}-\hat{\mathbf{P}}_{k})\bigr)\nonumber \\
		&\;+2\Re\bigl\{\Tr\bigl(\nabla_{\mathbf{X}_{k}}f(\hat{\mathbf{X}}_{k},\hat{\mathbf{P}}_{k})\herm(\mathbf{X}_{k}-\hat{\mathbf{X}}_{k})\bigr)\bigr\}.\label{eq:affapprox}
	\end{align}
	Note that we have used the fact that $\mathbf{P}_{k}\succeq0$ in deriving the above. To derive $\nabla_{\mathbf{X}_{k}}f(\mathbf{X}_{k},\mathbf{P}_{k})$ and $\nabla_{\mathbf{P}_{k}}f(\mathbf{X}_{k},\mathbf{P}_{k})$, we can rewrite $f(\mathbf{X}_{k},\mathbf{P}_{k})$ as 
	\[
	\log\det(\mathbf{P}_{k}-\mathbf{X}_{k}\mathbf{X}_{k}\herm)-\log\det(\mathbf{P}_{k}),
	\]
	which immediately gives
	\begin{equation}
		\nabla_{\mathbf{X}_{k}}f(\mathbf{X}_{k},\mathbf{P}_{k})=-(\mathbf{P}_{k}-\mathbf{X}_{k}\mathbf{X}_{k}\herm)^{-1}\mathbf{X}_{k}=-\mathbf{Y}_{k}^{-1}\mathbf{X}_{k}\label{eq:gradXk}
	\end{equation}
	and 
	
 \begin{align}
		\nabla_{\mathbf{P}_{k}}f(\mathbf{X}_{k},\mathbf{P}_{k})& =(\mathbf{P}_{k}-\mathbf{X}_{k}\mathbf{X}_{k}\herm)^{-1}-\mathbf{P}_{k}^{-1} \nonumber\\
		&=\mathbf{Y}_{k}^{-1}-(\mathbf{Y}_{k}+\mathbf{X}_{k}\mathbf{X}_{k}\herm)^{-1}.\label{eq:gradPk}
	\end{align}
	Using (\ref{eq:gradXk}) and (\ref{eq:gradPk}) in (\ref{eq:affapprox}) results in (\ref{eq:enequlityXY}) and thus completes the proof.
	
	\subsection{Proof of \thmref{closed-form-beamformer}\label{app:proof-precoder-closed-form}}	
	First, we can equivalently rewrite (\ref{eq:EneqRk1}) as
    \begin{align}
	g_{k}(\mathbf{F})&  \triangleq\log\det(\mathbf{I}+\hat{\mathbf{B}}_{k}\mathbf{\hat{\mathbf{X}}}_{k})-\Tr(\hat{\mathbf{B}}_{k}\mathbf{\hat{\mathbf{X}}}_{k})+\Tr(\hat{\mathbf{B}}_{k}\bar{\mathbf{H}}_{k}\mathbf{F}_{k})\nonumber \\
		&\hspace{-5pt}+\Tr((\bar{\mathbf{H}}_{k}\herm\hat{\mathbf{B}}_{k}\herm)\trans\mathbf{F}_{k}^{\ast})-\sum\nolimits_{j=1}^{K}\Tr((\bar{\mathbf{H}}_{k}\herm\hat{\mathbf{A}}_{k}\herm\bar{\mathbf{H}}_{k}\mathbf{F}_{j})\trans\mathbf{F}_{j}^{\ast}) \nonumber \\
		&\hspace{35pt}-\frac{\sigma_{0}^{2}}{P_{{\rm BS}}}\Tr(\hat{\mathbf{A}}_{k}\herm)\sum\nolimits_{i=1}^{K}\Tr((\bar{\mathbf{H}}\mathbf{F}_{i})\trans\mathbf{F}_{i}^{\ast}),\label{eq:EneqRk2}
	\end{align} where we have used the fact that $\Tr(\mathbf{X})=\Tr(\mathbf{X}\trans)$
	\cite[Eqn. (2.95)]{hjorungnes2011complex} and $\Tr(\mathbf{X}\mathbf{Y})=\Tr(\mathbf{Y}\trans\mathbf{X}\trans)$
	\cite[Eqn. (2.96)]{hjorungnes2011complex}. Next, using \cite[Table 3.2]{hjorungnes2011complex} we obtain
\begin{IEEEeqnarray}{c}
 \nabla_{\mathbf{F}_{j}}g_{k}(\mathbf{F}) =
\begin{cases}
\bar{\mathbf{H}}_{k}\herm\mathbf{B}_{k}\herm-\frac{\sigma_{0}^{2}}{P_{{\rm BS}}}\Tr(\hat{\mathbf{A}}_{k}\herm)\bar{\mathbf{H}}\mathbf{F}_{k}& {}\\
\hspace{40pt}-\bar{\mathbf{H}}_{k}\herm\hat{\mathbf{A}}_{k}\herm\bar{\mathbf{H}}_{k}\mathbf{F}_{k} & \hspace{-4pt}\text{if } j=k,
\\
\hspace{-2pt}-\frac{\sigma_{0}^{2}}{P_{{\rm BS}}}\Tr(\hat{\mathbf{A}}_{k}\herm)\bar{\mathbf{H}}\mathbf{F}_{j}-\bar{\mathbf{H}}_{k}\herm\hat{\mathbf{A}}_{k}\herm\bar{\mathbf{H}}_{k}\mathbf{F}_{j}  & \hspace{-4pt}\text{if } j\neq k.
\end{cases}\nonumber
\end{IEEEeqnarray}
Now, setting $\nabla_{\mathbf{F}_{j}}\sum_{k=1}^{K}\omega_{k}g_{k}(\mathbf{F})$ to zero yields
\begin{equation}
    \mathbf{F}_{j}=\bigl(\sum_{l=1}^{K}\frac{\omega_{l}\sigma_{0}^{2}}{P_{{\rm BS}}}\Tr(\hat{\mathbf{A}}_{l}\herm)\bar{\mathbf{H}}+\omega_{l}\bar{\mathbf{H}}_{l}\herm\hat{\mathbf{A}}_{l}\herm\bar{\mathbf{H}}_{l}\bigr)^{-1}\bar{\mathbf{H}}_{j}\herm\hat{\mathbf{B}}_{j}\herm\omega_{j},\forall j.\label{eq:closed-formFj}
\end{equation}
To finalize our proof, using $\mathbf{\tilde{A}}$, $\mathbf{\tilde{B}}$, and $\mu$ as defined in (\ref{eq:cloded_form_F}), we rewrite (\ref{eq:closed-formFj}) into more compact form as follows:
	\begin{subequations}
		\begin{flalign}
			\mathbf{F} & =(\mu\bar{\mathbf{H}}+\bar{\mathbf{H}}\mathbf{\tilde{A}}\bar{\mathbf{H}})^{-1}\bar{\mathbf{H}}\mathbf{\tilde{B}}.\label{eq:CF1}\\
			& =(\bar{\mathbf{H}}(\mu\mathbf{I}+\mathbf{\tilde{A}}\bar{\mathbf{H}}))^{-1}\bar{\mathbf{H}}\mathbf{\tilde{B}}\label{eq:CF2}\\
			& =(\mu\mathbf{I}+\mathbf{\tilde{A}}\bar{\mathbf{H}})^{-1}(\bar{\mathbf{H}})^{-1}\bar{\mathbf{H}}\mathbf{\tilde{B}}\label{eq:CF3}\\
			& =(\mu\mathbf{I}+\mathbf{\tilde{A}}\bar{\mathbf{H}})^{-1}\mathbf{\tilde{B}},\label{eq:CF4}
		\end{flalign}
	\end{subequations} where we have used the following assumptions: (\ref{eq:CF1}) is because
	$\bar{\mathbf{H}}=\bar{\mathbf{H}}\herm$; (\ref{eq:CF3}) follows the property of the inverse of product of matrices that if $\mathbf{U}$ and $\mathbf{V}$ are invertible then $\left(\mathbf{U}\mathbf{V}\right)^{-1}=\mathbf{V}^{-1}\mathbf{U}^{-1}$; and (\ref{eq:CF4}) assumes that $\mathbf{H}$ is a full row rank matrix, also employs the property of the invertible matrix that if
	$\mathbf{U}$ is invertible, then $\mathbf{U}^{-1}\mathbf{U}=\mathbf{I}$.
	
	\subsection{Proof of \thmref{SCA:convergence}\label{app:SCA:convergence:proof}}
	Suppose $\mathbf{F}^{(n)}$ is a nontrivial stationary solution, meaning that the corresponding sum rate is strictly positive. This implies that there exists some $m$ such that $||\mathbf{H}\herm\mathbf{F}_{m}^{(n)}||^{2}>0$. Otherwise, the achieved sum rate would be zero. At iteration $n+1$, it holds that	
     \begin{align}
        \hat{\mathbf{Y}}_{k}\hspace{-1pt} &\hspace{-2pt} =\hspace{-2pt}\sum_{j\neq k}^{K}\bar{\mathbf{H}}_{k}\mathbf{F}_{j}^{(n)}(\mathbf{F}_{j}^{(n)})\herm\bar{\mathbf{H}}_{k}\herm\hspace{-2pt}+\hspace{-2pt}\frac{\sigma_{0}^{2}}{P_{{\rm BS}}}\sum_{i=1}^{K}\Tr(\bar{\mathbf{H}}\mathbf{F}_{i}^{(n)}(\mathbf{F}_{i}^{(n)})\herm)\mathbf{I}\nonumber \\
          &\hspace{-2pt} =\sum_{j\neq k}^{K}\bar{\mathbf{H}}_{k}\mathbf{F}_{j}^{(n)}(\mathbf{F}_{j}^{(n)})\herm\bar{\mathbf{H}}_{k}\herm+\frac{\sigma_{0}^{2}}{P_{{\rm BS}}}\sum_{i=1,i\neq m}^{K}||\mathbf{H}\herm\mathbf{F}_{i}^{(n)}||^{2}\mathbf{I}\nonumber \\
         & \qquad+\frac{\sigma_{0}^{2}}{P_{{\rm BS}}}||\mathbf{H}\herm\mathbf{F}_{m}^{(n)}||^{2}\mathbf{I}\nonumber\\
         & \hspace{-2pt}\succeq\frac{\sigma_{0}^{2}}{P_{{\rm BS}}}||\mathbf{H}\herm\mathbf{F}_{m}^{(n)}||^{2}\mathbf{I}.
    \end{align}
	Consequently, $\hat{\mathbf{Y}}_{k}^{-1}$ exists and the lower bound $g_{k}(\mathbf{F})$ is well defined. If $||\mathbf{H}\herm\mathbf{F}_{k}^{(n)}||^{2}=0$, then $\tilde{R}_{k}(\mathbf{F})=0$. Conversely, if $||\mathbf{H}\herm\mathbf{F}_{k}^{(n)}||^{2}>0$, then it is straightforward to see that
	\begin{equation}
		\hat{\mathbf{Y}}_{k}\succeq\frac{\sigma_{0}^{2}}{P_{{\rm BS}}}||\mathbf{H}\herm\mathbf{F}_{k}^{(n)}||^{2}\mathbf{I}
	\end{equation}
	and thus 
	\begin{align}
		\hat{\mathbf{X}}_{k}\herm\mathbf{\hat{\mathbf{Y}}}_{k}^{-1}\hat{\mathbf{X}}_{k} & \preceq\frac{P_{{\rm BS}}}{\sigma_{0}^{2}||\mathbf{H}\herm\mathbf{F}_{k}^{(n)}||^{2}}\hat{\mathbf{X}}_{k}\herm\hat{\mathbf{X}}_{k}\\
		& \preceq\frac{P_{{\rm BS}}||\hat{\mathbf{X}}_{k}||^{2}}{\sigma_{0}^{2}||\mathbf{H}\herm\mathbf{F}_{k}^{(n)}||^{2}}\mathbf{I}
	\end{align}
	Clearly, $||\hat{\mathbf{X}}_{k}||^{2}=||\mathbf{H}_{k}\mathbf{H}\herm\mathbf{F}_{k}^{(n)}||^{2}\leq||\mathbf{H}_{k}||^{2}||\mathbf{H}\herm\mathbf{F}_{k}^{(n)}||^{2}$,
	leading to 
	\begin{equation}
		\hat{\mathbf{X}}_{k}\herm\mathbf{\hat{\mathbf{Y}}}_{k}^{-1}\hat{\mathbf{X}}_{k}\preceq\frac{P_{\rm BS}}{\sigma_{0}^{2}}||\mathbf{H}_{k}||^{2}\mathbf{I},
	\end{equation}
	which implies
        \begin{align}
        \tilde{R}_{k}(\mathbf{F}) & =\log\det((\mathbf{I}+\mathbf{X}_{k}\herm\mathbf{Y}_{k}^{-1}\mathbf{X}_{k})^{-1})\nonumber \\
         & \leq N_{r}\log(1+\frac{P_{{\rm BS}}}{\sigma_{0}^{2}}||\mathbf{H}_{k}||^{2}).
        \end{align}
Now that we have established that $\tilde{R}_{k}(\mathbf{F})$ is bounded above, its convergence is guaranteed because the sequence $\{\tilde{R}_{k}(\mathbf{F}^{(n)})\}$ is monotonically increasing according to the SCA principle. Let $\mathbf{F}^{\ast}$ be an accumulation point of the sequence $\{\mathbf{F}^{(n)}\}$. Then, the monotonic convergence of the sequence $\{\tilde{R}_{k}(\mathbf{F}^{(n)})\}$ indicates that it converges to $\tilde{R}_{k}(\mathbf{F}^{\ast})$. Following the same arguments in \cite[Chap. 2]{Bertsekas1999NP}, we can prove that $\mathbf{F}^{\ast}$ is a stationary solution of $(\mathcal{P}_{2})$.
	
	\subsection{Proof of \thmref{grad:theta}\label{app:proof-grad-theta-P1}}
	In this appendix, we provide a proof for the complex-valued gradient of (\ref{eq:rate-theta-givenW}) which is given by \thmref{grad:theta}. In particular, we will follow closely the steps detailed in \cite[Sec. 3.3.1]{hjorungnes2011complex}. Let $\mathbf{Z}_{k}=(\mathbf{H}_{k}\bar{\mathbf{W}}\mathbf{H}_{k}\herm+\sigma_{0}^{2}\mathbf{I})^{-1}$ and $\tilde{\mathbf{Z}}_{k} =(\mathbf{H}_{k}\bar{\mathbf{W}}_{k}\mathbf{H}_{k}\herm+\sigma_{0}^{2}\mathbf{I})^{-1}$. Then we have
	\begin{align}
		dR_{k}(\boldsymbol{\theta}) & =\Tr(\mathbf{Z}_{k}d(\mathbf{H}_{k})\bar{\mathbf{W}}\mathbf{H}_{k}\herm)+\Tr(\mathbf{Z}_{k}\mathbf{H}_{k}\bar{\mathbf{W}}d(\mathbf{H}_{k}\herm))\nonumber \\
		&\hspace{-15pt}-\Tr(\tilde{\mathbf{Z}}_{k}d(\mathbf{H}_{k})\bar{\mathbf{W}}_{k}\mathbf{H}_{k}\herm)-\Tr(\tilde{\mathbf{Z}}_{k}\mathbf{H}_{k}\bar{\mathbf{W}}_{k}d(\mathbf{H}_{k}\herm)),\label{eq:dfk1}
	\end{align}
	where we have used \cite[(3.60)]{hjorungnes2011complex} and \cite[Table 3.1]{hjorungnes2011complex}. It is easy to see that $d(\mathbf{H}_{k})=\mathbf{U}_{k}d(\boldsymbol{\mathbf{T}}(\boldsymbol{\theta}))\mathbf{G}$ and thus $d(\mathbf{H}_{k}\herm)=\mathbf{G}\herm d(\boldsymbol{\mathbf{T}}(\boldsymbol{\theta})\herm)\mathbf{U}_{k}\herm$. Subsequently, we can rewrite (\ref{eq:dfk1}) as
	\begin{align}
		dR_{k}(\boldsymbol{\theta}) & =\Tr(\mathbf{G}\bar{\mathbf{W}}\mathbf{H}_{k}\herm\mathbf{Z}_{k}\mathbf{U}_{k}d(\boldsymbol{\mathbf{T}}(\boldsymbol{\theta})))\nonumber \\
		&\quad+\Tr((\mathbf{U}_{k}\herm\mathbf{Z}_{k}\mathbf{H}_{k}\bar{\mathbf{W}}\mathbf{G}\herm)\trans d(\boldsymbol{\mathbf{T}}(\boldsymbol{\theta})^{\ast}))\nonumber \\
		&\quad-\Tr(\mathbf{G}\bar{\mathbf{W}}_{k}\mathbf{H}_{k}\herm\tilde{\mathbf{Z}}_{k}\mathbf{U}_{k}d(\boldsymbol{\mathbf{T}}(\boldsymbol{\theta})))\nonumber \\
		&\quad-\Tr((\mathbf{U}_{k}\herm\tilde{\mathbf{Z}}_{k}\mathbf{H}_{k}\bar{\mathbf{W}}_{k}\mathbf{G}\herm)\trans d(\boldsymbol{\mathbf{T}}(\boldsymbol{\theta})^{\ast})),\label{eq:dfk2}
	\end{align}
	where the properties of trace operator in \cite[Eqn. (2.95)]{hjorungnes2011complex} and \cite[Eqn. (2.96)]{hjorungnes2011complex} were applied. We note that $\boldsymbol{\mathbf{T}}(\boldsymbol{\theta})$ is a diagonal matrix and thus $dR_{k}(\boldsymbol{\theta})$ can be simplified to
	\begin{align}
		dR_{k}(\boldsymbol{\theta}) & =\vecd(\mathbf{G}\bar{\mathbf{W}}\mathbf{H}_{k}\herm\mathbf{Z}_{k}\mathbf{U}_{k})d(\boldsymbol{\theta})\nonumber \\
		&\quad+\vecd(\mathbf{U}_{k}\herm\mathbf{Z}_{k}\mathbf{H}_{k}\bar{\mathbf{W}}\mathbf{G}\herm)\trans d(\boldsymbol{\theta}^{\ast})\nonumber \\
		&\quad-\vecd(\mathbf{G}\bar{\mathbf{W}}_{k}\mathbf{H}_{k}\herm\tilde{\mathbf{Z}}_{k}\mathbf{U}_{k})d(\boldsymbol{\theta})\nonumber \\
		&\quad-\vecd(\mathbf{U}_{k}\herm\tilde{\mathbf{Z}}_{k}\mathbf{H}_{k}\bar{\mathbf{W}}_{k}\mathbf{G}\herm)\trans d(\boldsymbol{\theta}^{\ast}).\label{eq:dfk3}
	\end{align}
	Using \cite[Table 3.2]{hjorungnes2011complex} we obtain 
\begin{align}
		\nabla_{\boldsymbol{\theta}}R(\boldsymbol{\theta}) \hspace{-2pt}\triangleq\hspace{-2pt}\sum_{k=1}^{K}\omega_{k}\nabla_{\boldsymbol{\theta}}R_{k}(\boldsymbol{\theta})&\hspace{-2pt}=\hspace{-2pt}\sum_{k=1}^{K}\omega_{k} (\vecd(\mathbf{U}_{k}\herm\mathbf{Z}_{k}\mathbf{H}_{k}\bar{\mathbf{W}}\mathbf{G}\herm)\nonumber \\
		&\hspace{-10pt}-\vecd(\mathbf{U}_{k} \herm\tilde{\mathbf{Z}}_{k}\mathbf{H}_{k}\bar{\mathbf{W}}_{k}\mathbf{G}\herm)),
	\end{align}
which is indeed (\ref{eq:dRwtheta}) and thus completes our proof.
	
	\subsection{Complex-valued Gradient of $\tilde{R}(\boldsymbol{\theta})$ in (\ref{eq:rate-equivalent-theta})\label{app:rate-equivalent-grad}} 
	To support our argument in \subsecref{On-the-Numerical_P2}, we derive $\nabla_{\boldsymbol{\theta}}\tilde{R}(\boldsymbol{\theta})$ for a RIS-aided MISO system in this appendix. Let us first rewrite $\tilde{R}(\boldsymbol{\theta})$ in (\ref{eq:rate-equivalent-theta}) as
	\begin{align*}
		\tilde{R}(\boldsymbol{\theta}) =\sum\nolimits_{k=1}^{K}\omega_{k}\log\bigl(\sum_{j=1}^{K}|\bar{\mathbf{h}}_{k}\mathbf{f}_{j}|^{2}+\frac{\sigma_{0}^{2}}{P_{{\rm BS}}}\sum_{m=1}^{K}\mathbf{f}_{m}\herm\bar{\mathbf{H}}\mathbf{f}_{m}\bigr)\nonumber \\ -\sum\nolimits_{k=1}^{K}\omega_{k}\log\bigl(\sum\nolimits_{j\neq k}^{K}|\bar{\mathbf{h}}_{k}\mathbf{f}_{j}|^{2}+\frac{\sigma_{0}^{2}}{P_{{\rm BS}}}\sum\nolimits_{m=1}^{K}\mathbf{f}_{m}\herm\bar{\mathbf{H}}\mathbf{f}_{m}\bigr),
	\end{align*}
	where we have written $\mathbf{f}_{k}$ and $\bar{\mathbf{h}}_{k}$ instead of $\mathbf{F}_{k}$ and $\bar{\mathbf{H}}_{k}$, respectively, to emphasize that the channels and precoders now become vectors. Following the same steps in \appref{proof-grad-theta-P1} we have
	\begin{align}
		d\tilde{R}(\boldsymbol{\theta}) \hspace{-2pt}& =\hspace{-2pt}\sum\nolimits_{k=1}^{K}\hspace{-2pt}\omega_{k}\bigl(\sum\nolimits_{j=1}^{K}\hspace{-2pt}|\bar{\mathbf{h}}_{k}\mathbf{f}_{j}|^{2}\hspace{-2pt}+\hspace{-2pt}\frac{\sigma_{0}^{2}}{P_{{\rm BS}}}\sum\nolimits_{m=1}^{K}\mathbf{f}_{m}\herm\bar{\mathbf{H}}\mathbf{f}_{m}\bigr)^{-1}\nonumber \\
		& \hspace{-2pt}\times \bigl(\sum\nolimits_{j=1}^{K}d|\bar{\mathbf{h}}_{k}\mathbf{f}_{j}|^{2}+\frac{\sigma_{0}^{2}}{P_{{\rm BS}}}\sum\nolimits_{m=1}^{K}d(\mathbf{f}_{m}\herm\bar{\mathbf{H}}\mathbf{f}_{m})\bigr)\nonumber \\
		& \hspace{-12pt}-\sum\nolimits_{k=1}^{K}\omega_{k}\bigl(\sum\nolimits_{j\neq k}^{K}|\bar{\mathbf{h}}_{k}\mathbf{f}_{j}|^{2}+\frac{\sigma_{0}^{2}}{P_{{\rm BS}}}\sum\nolimits_{m=1}^{K}\mathbf{f}_{m}\herm\bar{\mathbf{H}}\mathbf{f}_{m}\bigr)^{-1}\nonumber \\
		&\times \bigl(\sum\nolimits_{j\neq k}^{K}d|\bar{\mathbf{h}}_{k}\mathbf{f}_{j}|^{2}+\frac{\sigma_{0}^{2}}{P_{{\rm BS}}}\sum\nolimits_{m=1}^{K}d(\mathbf{f}_{m}\herm\bar{\mathbf{H}}\mathbf{f}_{m})\bigr).\label{eq:dftheta_eq}
	\end{align}
	To obtain $d|\bar{\mathbf{h}}_{k}\mathbf{f}_{j}|^{2}$ we have the following equalities by
	\begin{subequations}
		\begin{align}
			d|\bar{\mathbf{h}}_{k}\mathbf{f}_{j}|^{2} & =d(\bar{\mathbf{h}}_{k})\mathbf{f}_{j}\mathbf{f}_{j}\herm\bar{\mathbf{h}}_{k}\herm+\bar{\mathbf{h}}_{k}\mathbf{f}_{j}\mathbf{f}_{j}\herm d(\bar{\mathbf{h}}_{k}\herm)\label{eq:dhkfj}\\
			d(\bar{\mathbf{h}}_{k}) & =d(\mathbf{h}_{k}\mathbf{H}^{H})=d(\mathbf{h}_{k})\mathbf{H}^{H}+\mathbf{h}_{k}d(\mathbf{H}^{H})\label{eq:dhkbar}\\
			d(\mathbf{h}_{k}) & =d(\mathbf{d}_{k}+\boldsymbol{\theta}\trans\mathbf{E}_{k})=d(\boldsymbol{\theta}\trans)\mathbf{E}_{k},\label{eq:dhk}
		\end{align}
	\end{subequations}
	where $\mathbf{E}_{k}=\diag\left(\mathbf{u}_{k}\right)\mathbf{G}.$ Next, it is easy to show that
	\begin{align}
		d(\mathbf{H}^{H}) &=\begin{bmatrix}\mathbf{E}_{k}\herm d(\boldsymbol{\theta}^{\ast}), & \ldots, & \mathbf{E}_{K}\herm d(\boldsymbol{\theta}^{\ast})\end{bmatrix}
	\end{align}
	and thus 
	\begin{align}
		d(\mathbf{H}) =\begin{bmatrix}d(\boldsymbol{\theta}\trans)\mathbf{E}_{1}\\
			\vdots\\
			d(\boldsymbol{\theta}\trans)\mathbf{E}_{K}
		\end{bmatrix}&=\begin{bmatrix}d(\boldsymbol{\theta}\trans)\\
			&   \ddots\\
			&  &  d(\boldsymbol{\theta}\trans)
		\end{bmatrix}\begin{bmatrix}\mathbf{E}_{1}\\
			\vdots\\
			\mathbf{E}_{K}
		\end{bmatrix}\nonumber \\
		&=\bigl(\mathbf{I}_{K}\otimes d(\boldsymbol{\theta}\trans)\bigr)\mathbf{E},\label{eq:dHherm}
	\end{align} where $\otimes$ denotes the Kronecker product and $\mathbf{E}=\begin{bmatrix}\mathbf{E}_{1}\trans & \mathbf{E}_{2}\trans & \ldots & \mathbf{E}_{K}\trans\end{bmatrix}\trans\in C^{(KN)\times M}.$ Inserting (\ref{eq:dhk}) and (\ref{eq:dHherm}) into (\ref{eq:dhkbar}) yields	
    \begin{align}
        d(\bar{\mathbf{h}}_{k}) &  =d(\boldsymbol{\theta}\trans)\mathbf{E}_{k}\mathbf{H}^{H}+\tilde{\mathbf{h}}_{k}\bigl(\mathbf{I}_{K}\otimes d(\boldsymbol{\theta}^{\ast})\bigr),
    \end{align}	where $\tilde{\mathbf{h}}_{k}=\mathbf{h}_{k}\mathbf{E}\herm\in\mathbb{C}^{1\times(KN)}$ and we have used the fact that $(\mathbf{A}\otimes\mathbf{B})\herm=\mathbf{A}\herm\otimes\mathbf{B}\herm$ \cite[Lemma 2.9]{hjorungnes2011complex}. Using the identity $\vect(\mathbf{A}\mathbf{B})=(\mathbf{I}_{m}\otimes\mathbf{A})\vect(\mathbf{B})$ for $\mathbf{A}$ and $\mathbf{B}$ of dimensions $k\times l$ and $l\times m$ \cite[Lemma 2.11]{hjorungnes2011complex}, respectively, we have
	\begin{equation}
		\vect\bigl(\tilde{\mathbf{h}}_{k}\bigl(I_{K}\otimes d(\boldsymbol{\theta}^{\ast})\bigr)\bigr)=\bigl(\mathbf{I}_{K}\otimes\tilde{\mathbf{h}}_{k}\bigr)\vect\bigl(\mathbf{I}_{K}\otimes d(\boldsymbol{\theta}^{\ast})\bigr).\label{eq:step1}
	\end{equation}
	Next using the identity $\vect(\mathbf{A}\otimes\mathbf{B})=(((\mathbf{I}_{n}\otimes\mathbf{K}_{qm})(\vect(\mathbf{A})\otimes\mathbf{I}_{q}))\otimes\mathbf{I}_{p})\vect(\mathbf{B})$ for $\mathbf{A}$ and $\mathbf{B}$ of dimensions $m\times n$ and $p\times q$, respectively where $\mathbf{K}_{qm}$ is the commutation	matrix of size $(qm)\times(qm)$ \cite[Lemma 2.13]{hjorungnes2011complex}, we have
	\begin{subequations}
		\begin{align}
			\vect(\mathbf{I}_{K}\otimes d(\boldsymbol{\theta})) & =(((\mathbf{I}_{K}\otimes\mathbf{K}_{1K})\vect(\mathbf{I}_{K}))\otimes\mathbf{I}_{N})d(\boldsymbol{\theta}^{\ast})\\
			& \hspace{-5pt}=(((\mathbf{I}_{K}\otimes\mathbf{I}_{K})\vect(\mathbf{I}_{K}))\otimes\mathbf{I}_{N})d(\boldsymbol{\theta}^{\ast})\\
			& \hspace{-5pt}=(\vect(\mathbf{I}_{K})\otimes\mathbf{I}_{N})d(\boldsymbol{\theta}^{\ast})\\
			& \hspace{-5pt}=\mathbf{P}d(\boldsymbol{\theta}^{\ast}),\label{eq:step2}
		\end{align}
	\end{subequations}
	where $\mathbf{P}=\vect(\mathbf{I}_{K})\otimes\mathbf{I}_{N}$. Thus we have 
	\begin{subequations}
		\begin{align}
			d(\bar{\mathbf{h}}_{k}) & =d(\boldsymbol{\theta}\trans)\mathbf{E}_{k}\mathbf{H}\herm+d(\boldsymbol{\theta}\herm)\mathbf{P}\trans(\mathbf{I}_{K}\otimes\tilde{\mathbf{h}}_{k}\trans)\label{eq:dhkbar-1}\\
			d(\bar{\mathbf{h}}_{k}\herm) & =\mathbf{H}\mathbf{E}_{k}\herm d(\boldsymbol{\theta}^{\ast})+(\mathbf{I}_{K}\otimes\tilde{\mathbf{h}}_{k}^{\ast})\mathbf{P}d(\boldsymbol{\theta}).\label{eq:dhkbar-2}
		\end{align}
	\end{subequations}
	Substituting (\ref{eq:dhkbar-1}) and (\ref{eq:dhkbar-2}) into (\ref{eq:dhkfj}) yields
\begin{align}
    d|\bar{\mathbf{h}}_{k}\mathbf{f}_{j}|^{2} & =\bigl(\mathbf{E}_{k}\mathbf{H}\herm\mathbf{f}_{j}\mathbf{f}_{j}\herm\bar{\mathbf{h}}_{k}\herm\bigr)\trans d(\boldsymbol{\theta})\nonumber\\
     & \quad+(\mathbf{P}\trans(\mathbf{I}_{K}\otimes\tilde{\mathbf{h}}_{k}\trans)\mathbf{f}_{j}\mathbf{f}_{j}\herm\bar{\mathbf{h}}_{k}\herm)\trans d(\boldsymbol{\theta}^{\ast})\nonumber\\
     & \quad+\bar{\mathbf{h}}_{k}\mathbf{f}_{j}\mathbf{f}_{j}\herm\mathbf{H}\mathbf{E}_{k}\herm d(\boldsymbol{\theta}^{\ast})\nonumber\\
     & \quad+\bar{\mathbf{h}}_{k}\mathbf{f}_{j}\mathbf{f}_{j}\herm(\mathbf{I}_{K}\otimes\tilde{\mathbf{h}}_{k}^{\ast})\mathbf{P}d(\boldsymbol{\theta}),
\end{align}
	which means
    \begin{align}
        \nabla_{\boldsymbol{\theta}}|\bar{\mathbf{h}}_{k}\mathbf{f}_{j}|^{2} & =\mathbf{P}\trans(\mathbf{I}_{K}\otimes\tilde{\mathbf{h}}_{k}\trans)\mathbf{f}_{j}\mathbf{f}_{j}\herm\bar{\mathbf{h}}_{k}\herm+\bigl(\bar{\mathbf{h}}_{k}\mathbf{f}_{j}\mathbf{f}_{j}\herm\mathbf{H}\mathbf{E}_{k}\herm\bigr)\trans.\label{eq:dhkfj2}
    \end{align}
	Similarly, we have
    \begin{subequations}
        \begin{align}
        d(\mathbf{f}_{m}\herm\bar{\mathbf{H}}\mathbf{f}_{m}) & =\mathbf{f}_{m}\herm d(\bar{\mathbf{H}})\mathbf{f}_{m}\\
         & =\mathbf{f}_{m}\herm d(\mathbf{H})\mathbf{H}\herm\mathbf{f}_{m}+\mathbf{f}_{m}\herm\mathbf{H}d(\mathbf{H}\herm)\mathbf{f}_{m}\\
         & =\mathbf{f}_{m}\herm\bigl(\mathbf{I}_{K}\otimes d(\boldsymbol{\theta}\trans)\bigr)\mathbf{E}\mathbf{H}\herm\mathbf{f}_{m}\nonumber \\
         & \qquad+\mathbf{f}_{m}\herm\mathbf{H}\mathbf{E}\herm\bigl(\mathbf{I}_{K}\otimes d(\boldsymbol{\theta}^{\ast})\bigr)\mathbf{f}_{m}.\label{eq:dfmHfm}
        \end{align}
    \end{subequations}
	Using the steps from (\ref{eq:step1}) to (\ref{eq:step2}) we can prove that
	\begin{subequations}
		\begin{align}
			\mathbf{f}_{m}\herm\bigl(\mathbf{I}_{K}\otimes d(\boldsymbol{\theta}\trans)\bigr) & =d(\boldsymbol{\theta}\trans)\mathbf{P}\trans\bigl(\mathbf{f}_{m}^{\ast}\otimes\mathbf{I}_{KN}\bigr)\label{eq:dstep1}\\
			\bigl(\mathbf{I}_{K}\otimes d(\boldsymbol{\theta}^{\ast})\bigr)\mathbf{f}_{m} & =\bigl(\mathbf{f}_{m}\trans\otimes\mathbf{I}_{KN}\bigr)\mathbf{P}d(\boldsymbol{\theta}^{\ast}).\label{eq:dstep2}
		\end{align}
	\end{subequations}
	Substituting (\ref{eq:dstep1}) and (\ref{eq:dstep2}) into (\ref{eq:dfmHfm}) gives
	\begin{align}	d(\mathbf{f}_{m}\herm\bar{\mathbf{H}}\mathbf{f}_{m})=&d(\boldsymbol{\theta}\trans)\mathbf{P}\trans\bigl(\mathbf{f}_{m}^{\ast}\otimes\mathbf{I}_{KN}\bigr)\mathbf{E}\mathbf{H}\herm\mathbf{f}_{m}\nonumber \\
		&\quad+\mathbf{f}_{m}\herm\mathbf{H}\mathbf{E}\herm\bigl(\mathbf{f}_{m}\trans\otimes\mathbf{I}_{KN}\bigr)\mathbf{P}d(\boldsymbol{\theta}^{\ast}),
	\end{align}
        and thus,
	\begin{equation}
		\nabla_{\boldsymbol{\theta}}(\mathbf{f}_{m}\herm\bar{\mathbf{H}}\mathbf{f}_{m})=\mathbf{P}\trans(\mathbf{f}_{m}\otimes\mathbf{I}_{KN})\mathbf{E}^{\ast}\mathbf{H}\trans\mathbf{f}_{m}^{\ast}.\label{eq:dfmHbarfm}
	\end{equation}
	By substituting (\ref{eq:dhkfj2}) and (\ref{eq:dfmHbarfm}) into	(\ref{eq:dftheta_eq}) we obtain $\nabla_{\boldsymbol{\theta}}\tilde{R}(\boldsymbol{\theta})$ given in (\ref{eq:rate-equivalent-grad}) at the top of the next page.
    \begin{figure*}[!t]
		\begin{align}
		\nabla_{\boldsymbol{\theta}}\tilde{R}(\boldsymbol{\theta}) & =\sum\nolimits_{k=1}^{K}\omega_{k}\left(\frac{\sum_{j=1}^{K}\bigl(\mathbf{P}\trans\bigl(\mathbf{I}_{K}\otimes\tilde{\mathbf{h}}_{k}\trans\bigr)\mathbf{f}_{j}\mathbf{f}_{j}\herm\bar{\mathbf{h}}_{k}\herm+\bigl(\bar{\mathbf{h}}_{k}\mathbf{f}_{j}\mathbf{f}_{j}\herm\mathbf{H}\mathbf{E}_{k}\herm\bigr)\trans\bigr)+\frac{\sigma_{0}^{2}}{P_{{\rm BS}}}\sum_{m=1}^{K}\mathbf{P}\trans\bigl(\mathbf{f}_{m}\otimes\mathbf{I}_{KN}\bigr)\mathbf{E}^{\ast}\mathbf{H}\trans\mathbf{f}_{m}^{\ast}}{\sum_{k=1}^{K}|\bar{\mathbf{h}}_{k}\mathbf{f}_{k}|^{2}+\frac{\sigma_{0}^{2}}{P_{{\rm BS}}}\sum_{m=1}^{K}\mathbf{f}_{m}\herm\bar{\mathbf{H}}\mathbf{f}_{m}}\right)\nonumber \\
			& -\sum\nolimits_{k=1}^{K}\omega_{k}\left(\frac{\sum_{j\neq k}^{K}\bigl(\mathbf{P}\trans\bigl(\mathbf{I}_{K}\otimes\tilde{\mathbf{h}}_{k}\trans\bigr)\mathbf{f}_{j}\mathbf{f}_{j}\herm\bar{\mathbf{h}}_{k}\herm+\bigl(\bar{\mathbf{h}}_{k}\mathbf{f}_{j}\mathbf{f}_{j}\herm\mathbf{H}\mathbf{E}_{k}\herm\bigr)\trans\bigr)+\frac{\sigma_{0}^{2}}{P_{{\rm BS}}}\sum_{m=1}^{K}\mathbf{P}\trans\bigl(\mathbf{f}_{m}\otimes\mathbf{I}_{KN}\bigr)\mathbf{E}^{\ast}\mathbf{H}\trans\mathbf{f}_{m}^{\ast}}{\sum_{j\neq k}^{K}|\bar{\mathbf{h}}_{k}\mathbf{f}_{j}|^{2}+\frac{\sigma_{0}^{2}}{P_{{\rm BS}}}\sum_{m=1}^{K}\mathbf{f}_{m}\herm\bar{\mathbf{H}}\mathbf{f}_{m}}\right)\label{eq:rate-equivalent-grad}
		\end{align}
		\hrulefill{}
	\end{figure*} 
    \subsection{Proof of \lemref{linesearch}\label{app:proof-LS}}
    In this appendix, we prove that the proposed line search in \algref{AO} terminates after a finite number of steps. Since a scaled projected method is used in our proposed algorithm, it is more convenient to consider a transform of variables defined by
	\begin{equation}
		\bar{\boldsymbol{\theta}}=\boldsymbol{\boldsymbol{\Xi}}^{-1/2}\boldsymbol{\theta}.
	\end{equation}
	Let $h(\bar{\boldsymbol{\theta}})\triangleq R(\boldsymbol{\boldsymbol{\Xi}}^{1/2}\bar{\boldsymbol{\theta}})$.
	Then it follows immediately that $\nabla_{\bar{\boldsymbol{\theta}}}h(\bar{\boldsymbol{\theta}}) =\boldsymbol{\boldsymbol{\Xi}}^{1/2}\nabla_{\boldsymbol{\theta}}R(\boldsymbol{\theta})$ and that 
    \begin{equation}		\boldsymbol{\theta}^{(\ell)}+\alpha_{\ell+1}\boldsymbol{\boldsymbol{\Xi}}\nabla_{\boldsymbol{\theta}}R(\boldsymbol{\theta}^{(\ell)})=\boldsymbol{\boldsymbol{\Xi}}^{1/2}\bigl(\bar{\boldsymbol{\theta}}^{(\ell)}+\alpha_{\ell+1}\nabla_{\bar{\boldsymbol{\theta}}}h(\bar{\boldsymbol{\theta}}^{(\ell)})\bigr).
\end{equation}
	Thus, in the space of $\bar{\boldsymbol{\theta}}$, the projected step in (\ref{eq:mPG}) is equivalent to
    \begin{equation}		\bar{\boldsymbol{\theta}}^{(\ell+1)}=\Pi_{\bar{\mathcal{Q}}}\bigl(\bar{\boldsymbol{\theta}}^{(\ell)}+\alpha_{\ell+1}\nabla_{\bar{\boldsymbol{\theta}}}h(\bar{\boldsymbol{\theta}}^{(\ell)})\bigr),
    \end{equation}
	where $\bar{\mathcal{Q}}=\{\bar{\boldsymbol{\theta}}=\boldsymbol{\boldsymbol{\Xi}}^{-1/2}\boldsymbol{\theta}\big|\boldsymbol{\theta}\in\mathcal{Q}\}$. The projection is equivalent to
	\begin{align}
        \bar{\boldsymbol{\theta}}^{(\ell+1)} & =\underset{\bar{\boldsymbol{\theta}}\in\bar{\mathcal{Q}}}{\argmin}\ \Vert\bar{\boldsymbol{\theta}}-\bigl(\bar{\boldsymbol{\theta}}^{(\ell)}+\alpha_{\ell+1}\nabla_{\bar{\boldsymbol{\theta}}}h(\bar{\boldsymbol{\theta}}^{(\ell)})\bigr)\Vert\nonumber \\
         & =\underset{\bar{\boldsymbol{\theta}}\in\bar{\mathcal{Q}}}{\argmin}\ \frac{1}{2\alpha_{\ell+1}}\Vert\bar{\boldsymbol{\theta}}-\bar{\boldsymbol{\theta}}^{(\ell)}\Vert\nonumber \\
         & \qquad\qquad\qquad-2\Re\{\nabla_{\bar{\boldsymbol{\theta}}}h(\bar{\boldsymbol{\theta}}^{(\ell)})\herm(\bar{\boldsymbol{\theta}}-\bar{\boldsymbol{\theta}}^{(\ell)})\},
        \end{align}
	which gives
	\begin{equation}
		\frac{1}{2\alpha_{\ell+1}}\Vert\bar{\boldsymbol{\theta}}^{(\ell+1)}-\bar{\boldsymbol{\theta}}^{(\ell)}\Vert-2\Re\{\nabla_{\bar{\boldsymbol{\theta}}}h(\bar{\boldsymbol{\theta}}^{(\ell)})\herm(\bar{\boldsymbol{\theta}}^{(\ell+1)}-\bar{\boldsymbol{\theta}}^{(\ell)})\})\leq0.\label{eq:projected-step}
	\end{equation}
	Let $L$ be the Lipschitz constant of $\nabla_{\bar{\boldsymbol{\theta}}}h(\bar{\boldsymbol{\theta}})$. Then we have 
	\begin{align}
		h(\bar{\boldsymbol{\theta}}^{(\ell+1)})\geq h(\bar{\boldsymbol{\theta}}^{(\ell)})&+2\Re\{\nabla_{\bar{\boldsymbol{\theta}}}h(\bar{\boldsymbol{\theta}}^{(\ell)})\herm(\bar{\boldsymbol{\theta}}^{(\ell+1)}-\bar{\boldsymbol{\theta}}^{(\ell)})\}\nonumber \\
		&-\frac{L}{2}\Vert\bar{\boldsymbol{\theta}}^{(\ell+1)}-\bar{\boldsymbol{\theta}}^{(\ell)}\Vert. \label{eq:Lipschitzbound}
	\end{align}
	Combining (\ref{eq:projected-step}) and (\ref{eq:Lipschitzbound}) results in
	\begin{equation}
		h(\bar{\boldsymbol{\theta}}^{(\ell+1)})\geq h(\bar{\boldsymbol{\theta}}^{(\ell)})-\frac{1}{2}(L-\frac{1}{\alpha_{\ell+1}})\Vert\bar{\boldsymbol{\theta}}^{(\ell+1)}-\bar{\boldsymbol{\theta}}^{(\ell)}\Vert
	\end{equation}
	which is equivalent to
	\begin{equation}
		R(\boldsymbol{\theta}^{(\ell+1)})\geq R(\boldsymbol{\theta}^{(\ell)})-\frac{1}{2}(L-\frac{1}{\alpha_{\ell+1}})\Vert\boldsymbol{\boldsymbol{\Xi}}^{-1/2}(\boldsymbol{\theta}^{(\ell+1)}-\boldsymbol{\theta}^{(\ell)})\Vert.
	\end{equation}
	For a given $\beta$, it is obvious that there exists a sufficiently small $\frac{1}{\alpha_{\ell+1}}$ such that $-\frac{1}{2}(L-\frac{1}{\alpha_{\ell+1}})\Vert\boldsymbol{\boldsymbol{\Xi}}^{-1/2}(\boldsymbol{\theta}^{(\ell+1)}-\boldsymbol{\theta}^{(\ell)})\Vert\geq\frac{\beta}{2N_{s}}\Vert(\boldsymbol{\theta}^{(\ell+1)}-\boldsymbol{\theta}^{(\ell)})\Vert$.	
    \bibliographystyle{IEEEtran}
    \bibliography{IEEEabrv,References} 
\end{document}